\documentclass[12pt]{amsart}
\usepackage{amsthm}
\theoremstyle{definition}

\usepackage{mathrsfs}
\usepackage{amsmath}
\usepackage{bm}
\usepackage{tikz,xcolor}
\usepackage{float}
\usepackage{hyperref}
\usepackage{amssymb}
\usepackage{bbm}
\usepackage[round]{natbib}
\usepackage{footnote}
\makesavenoteenv{tabular}
\usepackage[titletoc]{appendix}
\usepackage[capitalise]{cleveref}
\usepackage{subcaption}
\usepackage{mathtools}
\usepackage{multirow}
\usepackage{comment}
\usepackage[margin=1in]{geometry}
\usepackage[normalem]{ulem}

\newcommand{\Xone}{\bm{X}^*}
\newcommand{\X}{\bm{X}}
\newcommand{\Pperpx}{{\bm{P}_X}^{\perp}}
\newcommand{\Pperpt}{{\bm{P}_T}^{\perp}}
\newcommand{\Px}{\bm{P}_X}
\newcommand{\Y}{\bm{Y}}
\newcommand{\W}{\bm{W}}
\newcommand{\F}{\bm{F}}

\newcommand{\Z}{\bm{Z}}
\newcommand{\s}{\tau_{s}}
\newcommand{\e}{\tau_{\epsilon}}
\newcommand{\betas}{\bm{\beta}^X}

\newcommand{\betass}[1]{\bm{\beta}_{#1}^X}

\newcommand{\betashat}[1]{\bm{\hat{\beta}}_{#1}^*}
\newcommand{\betax}{\beta^X}
\newcommand{\bs}{b_{\epsilon}}
\newcommand{\bt}{b_{s}}
\newcommand{\as}{a_{\epsilon}}
\newcommand{\at}{a_{s}}

\newtheorem*{remark*}{Remark}
\newtheorem*{theorem*}{Theorem}
\newtheorem{thm}{Theorem}
\newtheorem{lemma}{Lemma}
\newtheorem{cor}{Corollary}
\newtheorem{defn}{Definition}

\title[Restricted Spatial Regression Methods]{Restricted Spatial Regression Methods: \\ Implications for Inference}

\begin{document}
	
	\maketitle
	
\begin{center}
Kori Khan\footnote{Email: khan.746@osu.edu}, Department of Statistics, The Ohio State University\\ and\\ Catherine A. Calder, Department of Statistics and Data Sciences,\\University of Texas at Austin\\[.3in]

\end{center}
	
\begin{abstract}
	The issue of spatial confounding between the spatial random effect and the fixed effects in regression analyses has been identified as a concern in the statistical literature. Multiple authors have offered perspectives and potential solutions. In this paper, for the areal spatial data setting, we show that many of the methods designed to alleviate spatial confounding can be viewed as special cases of a general class of models. We refer to this class as Restricted Spatial Regression (RSR) models, extending terminology currently in use.  We offer a mathematically based exploration of the impact that RSR methods have on inference for regression coefficients for the linear model. We then explore whether these results hold in the generalized linear model setting for count data using simulations. We show that the use of these methods have counterintuitive consequences which defy the general expectations in the literature. In particular, our results and the accompanying simulations suggest that RSR methods will typically perform worse than non-spatial methods. These results have important implications for dimension reduction strategies in spatial regression modeling.   Specifically, we demonstrate that the problems with RSR models cannot be fixed with a selection of ``better" spatial basis vectors or dimension reduction techniques.
	
    \noindent {{\sc Keywords:} Confounding; Spatial Statistics; Bayesian; Dimension Reduction}
\end{abstract}

\section{Introduction}

In our increasingly data rich world, large spatial  data sets are becoming abundant.  A booming field of research involves incorporating spatial dependence into models in a computationally efficient way.  As is often the case with spatial statistics, much of this research has initially  focused on developing and analyzing methods for spatial process models: models developed to make predictions at unobserved locations \citep[e.g.,][]{banerjee2008gaussian, wikle2010low, fuentes2007approximate,stein2014limitations}. There have been few attempts to understand how these methods impact inference on regression coefficients.

Recently, however, there has been a line of work designed to efficiently incorporate spatial dependence into regression models when the primary interest is on inference for the regression coefficients.  In many ways, this set of work mirrors the work done in the spatial process model research. For some examples, in the context of areal data, \citet{Hughes} presented a reduced-rank approach, \citet{Prates} developed a sparse approximation technique, and \citet{burden2015sar} introduced an approximate likelihood method. \citet{bradley2015} extended ideas of basis selection techniques to multivariate spatio-temporal mixed effects models, while \citet{murakami2015} incorporate spatial dependence through eigen-vector spatial filtering.  More recently, \citet{Thaden} introduced a structural equation approach for estimating regression coefficients when there is an unobserved spatially dependent covariate.  

In contrast to the work in the spatial process model realm, these methods are either designed or inspired by methods designed to alleviate ``spatial confounding." The first explicit reference to spatial confounding is often attributed to \citet{Clayton}, who observed what they referred to as ``confounding by location": the situation where estimates of a regression coefficient associated with a spatially-structured covariate are affected by the presence of a spatial random effect in the model.  In recent work, the phenomenon is often explained as the presence of multicollinearity between the covariates, $\bm{X}$, and the spatial random effect \citep{Prates,Hanks,Thaden,Hefley}.

\citet{Reich} and \citet{Hodges2} brought recent attention to the issue  \citep[see also,][for a study on the effects of spatial confounding on inference for regression coefficients]{Pactwo}. These works highlighted for the first time that in the presence of spatial confounding, not only can the estimates for regression coefficients change, but the uncertainty associated with these estimates can be ``overinflated." To address these dual concerns, \citet{Reich} proposed a method employing synthetic predictors to smooth orthogonally to the fixed effects.  \citet{Hughes} then extended this work by suggesting the use of eigenvectors of the Moran operator as synthetic predictors. They argued these basis functions would allow for dimension reduction through the selection of only those synthetic predictors associated with ``attractive" spatial dependence (as opposed to ``repulsive" spatial dependence). The intuitive appeal of this set of work inspired a series of follow-up investigations \citep{Prates, burden2015sar,bradley2015,Thaden}. However, it has been observed that these methods can lead to elevated levels of Type-S errors, the Bayesian analogue of Type I error \citep[][the latter paper suggested a posterior predictive approach to address this concern]{Prates,Hanks}. Recently, \citet{Hanks} and \citet{Prates} have noted that there is a need to better understand when it is appropriate to utilize such methods. Indeed, with the exception of \citet{Reich}, there has been a lack of mathematical formalism to assess the impact of smoothing orthogonally to the fixed effects has on inference for regression coefficients. 

In this work, we propose to fill that void. To do so, we consider a Bayesian analysis of Gaussian areal spatial data. We show that with respect to inference on regression coefficients, the current methods proposed to smooth orthogonally to the fixed effects can be thought of as a subset of a larger class of models. Extending the terminology currently in use for these proposed methods, we will refer to this larger class as \textit{Restricted Spatial Regression} (RSR) models. We find that RSR models transform a mixed-effects model into an over-fit linear model. Specifically, any of these models will produce a posterior mean for the regression coefficients which is equivalent to the posterior mean obtained in the corresponding non-spatial model. The various approaches to smoothing orthogonally to fixed effects were designed to ensure that all the marginal posterior variances of the regression coefficients are greater than the corresponding posterior variances of the non-spatial model. However, we show that the exact opposite is often true. The reduction in the posterior variances is caused solely by the addition of the spatial basis functions.  Furthermore, our analytic results and the included simulation studies indicate that sufficiently large credible intervals for the regression coefficients will generally be nested within the corresponding credible intervals from the non-spatial model. Importantly, these results are invariant to the spatial structure of the covariates and any spatial structure in the residuals. In short, our results indicate that with respect to coverage and Type-S error, one would be better off fitting a non-spatial model than any RSR model. Furthermore, simulations indicate that when there is spatial dependence unexplained by the covariates, RSR models for Gaussian data exasperate the decrease in coverage and increase in Type-S rates - even if the true unexplained spatial dependence in the response variable is generated orthogonally to the covariates. A simulation study and example application of RSR models suggest similar results might hold for count data. 

The rest of the paper proceeds as follows: \cref{sec2} provides an overview of deriving inference on regression coefficients in a spatial random effects model and a discussion of the methods developed to alleviate spatial confounding. \cref{mywork} contains the results of this paper. \cref{simu} includes simulation studies, and \cref{example} compares various approaches on the Slovenia stomach cancer data set. The proofs for all theorems are contained in Appendix B.

\section{Background} \label{sec2}

\subsection{Inference on Regression Coefficients in Random Effects Models}
\sloppypar{
The spatial generalized linear mixed model (SGLMM), popularized by \citet{Diggle}, assumes that $\Y = \{ y_i , \ldots, y_n \}^T$ is a realization from a random field where $y_i$ is observed at spatial location $s_i, ~ i= 1 \ldots n$.  The vector of transformed conditional means $\bm{Z} = \{g(E(y_1 | \eta(s_1))), \ldots, g(E(y_n | \eta(s_n))\}^T$ for a given link function $g$ is then related to fixed effects and a spatial random effect:}
\begin{equation}
 \Z = \bm{1} \beta_0 +  \bm{X} \betas + \bm{\eta}. \label{SGLMM}
\end{equation}
Here, $\bm{1}$ is the $n \times 1$ column vector of 1's, $\X = [\X_1, \ldots, \X_p ]$ is the $n \times p$ design matrix whose $i$th row consists of the $p$ covariates associated with $\Z_i$ , $\bm{\beta}^* = (\beta_0, ~ \betas) \in \mathbb{R}^{p+1}$ is a vector of regression parameters, and $\bm{\eta} = (\eta_1, \ldots, \eta_n)^T$ is a zero-mean random effect with a spatial covariance matrix $\bm{\Sigma}(\bm{\theta})$ parameterized by $\bm{\theta} \in \mathbb{R}^m$.  In this paper, the notation denoting dependence on location will be dropped when the meaning is unambiguous. We define ${\X}^* \equiv [\bm{1} ~ \X  ]$. The distinction between ${\X}$ and ${\X}^*$ is important when we compare different models. However, unless otherwise stated, the points about one in this section are true for both. 

Historically, much of the statistical community's intuition for the behavior of estimates of regression coefficients is based on work for the linear model when $g$ is the identity mapping.  A non-spatial (NS) analysis in this case would correspond to an assumption that $\bm{\Sigma}(\theta) \propto \bm{I}$. In this setting, the best linear unbiased estimator for the regression coefficients is the well-known ordinary least squares (OLS) estimator $\betashat{NS} = ( {{\X}^*}^T {\X}^*)^{-1} {{\X}^*}^T \Y$. Classical statistics textbooks remind us that if the assumption of i.i.d error structure is not met, these OLS estimates will be inefficient. Instead, if $\bm{\theta}$ were known, the most efficient estimator would be the generalized least squares estimator (GLS) $\betashat{GLS} = ({\Xone}^T \bm{\Sigma}^{-1}(\bm{\theta}) \Xone)^{-1} {\Xone}^T \bm{\Sigma}(\bm{\theta})^{-1} \bm{Y}$.

The two estimates will generally not be the same. However, in spatial statistics there is often an assumption that the addition of a spatial random effect should not change the point estimates of the regression coefficients. This expectation seems to be driven by the fact that geostatistical software developed to facilitate spatial process modeling often only implemented ordinary kriging and not universal kriging, the former allowing for an unknown, but constant mean and the latter allowing the mean to be an unknown linear combination of covariates associated with points in space \citep[pg. 344]{Waller}. Conventional wisdom in the spatial statistics literature suggests that if there is spatial dependence unexplained by the covariates (i.e,  processes in which things near in space are assumed to be more similar than things far away in space), the naive use of $\betashat{NS}$ will underestimate the variance of this estimator. This intuition is driven by well-understood examples which illustrate that ignoring spatial dependence will result in an underestimate of the variance of the mean of a spatial process \citep[see, e.g.,][pages 13-15]{Cressie93}.  \citet{Pactwo} pointed out that there is a lack of formal quantification of this belief for regression coefficients in general. Working in a setting where $\bm{X}$ is stochastic, \citet{Pactwo} showed that generally the naive variance estimator $\textrm{Var}(\hat{\bm{{\beta}}}^X_{OLS} ) $ will underestimate the uncertainty estimate of the correct variance estimator $\textrm{Var}(\hat{\bm{{\beta}}}^X_{GLS} )$, thereby lending support to the common expectation in spatial statistics. In any case, the expectation that the estimates of the variance associated with regression coefficient estimators will increase in models accounting for the presence of residual spatial dependence extends beyond the linear model.  In practice this expectation is often true, but there are examples in the literature where it does not hold \citep[for an example, see][]{banerjee2003frailty}.

\subsection{Spatial Confounding}
The term spatial confounding is used to describe the effect that multicollinearity between the fixed covariates $\bm{X}$ and the spatial random effect $\bm{\eta}$ has on the point estimates and variance of the regression coefficient estimates. Most of the work in spatial confounding involves a fully Bayesian analysis where the point estimates for the regression coefficients are defined to be the posterior mean $E(\betas | \Y)$ and the variances of interest are the diagonal entries of $\textrm{Var} (\betas | \Y)$ \citep{Reich,Hughes,Hefley,Hanks,Prates}. Efforts have been made to propose statistics designed to detect the presence of spatial confounding \citep{Reich,Hefley,Prates}. \citet{Thaden} provided a formalization of spatial confounding that assumes the multicollinearity between $\X$ and the spatial random effect occurs because of the absence of another unobserved spatially varying covariate. However, there is not currently a formal definition of spatial confounding in a more general setting, in part because it is a difficult concept to quantify. It is easier to define what spatial confounding is not, and so that is what we do in  \cref{def:spatconf}. As will be seen shortly, all the methods designed to address spatial confounding are designed in the hopes of achieving the properties in \cref{def:spatconf}.

\begin{defn} \label{def:spatconf}
	A method which results in posterior mean $E(\betas | \Y)$ and marginal posterior variances $\textrm{Var}(\betas_i | \Y)$, $i =1 , \ldots, p$ alleviates spatial confounding if the following conditions are met:
	\begin{enumerate}
		\item $E(\betas | \Y) = E(\betass{NS} | \Y)$, and
		\item $ \textrm{Var}({\betass{NS,}}_i | \Y) \leq  \textrm{Var}({\bm{\beta}}_i^X | \Y) \leq  \textrm{Var}({\betass{Spatial,}}_i | \Y) $ for $i= 1, \ldots, p$,
	\end{enumerate}
where $ \betass{NS}$ are the regression coefficients of the corresponding non-spatial model and $\betass{Spatial}$ are the regression coefficients from an unrestricted spatial random effect.
\end{defn}

\citet{Hodges2} argued that spatial confounding can be a concern whenever a spatial random effect is included in a model. However, most of the proposed methods to address spatial confounding are developed in the context of Gaussian areal spatial data \citep{Reich,Hodges2,Hughes,Prates}. In the areal data setting, spatial dependence is described by the introduction of an underlying, undirected graph $G = (V, E)$. Non-overlapping spatial regions that partition the study area are represented by vertices, $V = \{ 1, \ldots, n\}$, and edges $E$ defined so that each pair $(i,j)$ represents the proximity between region $i$ and region $j$. We represent $G$ by its $n \times n$ binary adjacency matrix $\bm{A}$ with entries defined such that $\textrm{diag}(\bm{A}) = 0$ and $\bm{A}_{i,j} = \mathbbm{1}_{(i,j) \in E, i \neq j}$. In models designed to address spatial confounding, the spatial random effect is typically assumed to follow the intrinsic conditional autoregressive (ICAR) model/prior \citep{Besag}. We will refer to models utilizing this prior as ICAR models. 
 
For Gaussian areal data, we observe that the ICAR model, the non-spatial model, and models designed to alleviate spatial confounding such as those proposed by \citet{Reich} (RHZ), \citet{Hughes} (HH), and \citet{Prates} (PAR) are all special cases of the following more general form:
\begin{eqnarray}\label{second}
\Y = \bm{X} \betas + \bm{W} \bm{\delta}  + \bm{\epsilon} \\
p( \bm{\delta}| \tau_s )  \propto \tau_s^{\textrm{rank}(\F) /2} \exp \{ \frac{- \tau_s}{2} \bm{\delta}^T \bm{F} \bm{\delta} \nonumber\},
\end{eqnarray}
where $\bm{\epsilon} \sim N(\bm{0},\e \bm{I})$, and $\e$ and $\s$ are precision parameters. $\bm{W}$ is a $n \times q$ set of basis vectors such that $q \leq n-p$, and $\F$ is a symmetric, non-negative definite matrix.  \cref{tab1} illustrates how the three proposed RSR methods can be considered special cases of the form \eqref{second}. 

\begin{center}
	\captionsetup{position=above}
	\captionof{table}{Special Cases of \eqref{second}} \label{tab1}
	\begin{tabular}{ |p{2cm}||p{2cm}|p{2cm}|p{2cm}|  }
		\hline
		Model & \scriptsize Design Matrix  &$\bm{W}$ & $\bm{F}$ \\
		\hline
		NS & $\Xone$    & $\bm{0}$ &   $\bm{0}$ \\
		ICAR   & $\X$    & $\bm{I}$ &   $ \bm{Q}$ \\
		RHZ  & $\Xone$    & $\bm{L}$ &   $\bm{L}^T \bm{Q} \bm{L}$ \\
		HH & $\Xone$  & $\bm{M}_q$ &  $\bm{M}_q^T \bm{Q} \bm{M}_q$ \\
		PAR  & $\X$ & $\bm{I}$ & $\bm{Q}^{\perp}$ \\ %\footnote{Even though $\bm{Q}$ is connected, there is no need for $\bm{Q}^{\perp}$ to be connected. For simplicity, we assume for the moment that it is.}   \\
		\hline 
	\end{tabular}\\[.3in]
\end{center}

The distinction between the use of $\X$ and $\X^*$ in \cref{tab1} is a direct consequence of the impropriety of the ICAR prior. The ICAR model captures spatial dependence through a Gaussian Markov Random Field (GMRF). The precision matrix is the graph Laplacian $\bm{Q} = \textrm{diag}(\bm{A} \bm{1}) - \bm{A}$. Because the adjacency matrix $\bm{A}$ is defined to be the zero/one adjacency matrix on $G$, $\bm{Q}$ is the graph Laplacian of a simple graph. A well-known fact from spectral graph theory is that the kernel of $\bm{Q}$ will be constant on each connected component of $G$ \citep{von2007tutorial}.  This fact means that the prior $p(\bm{\delta} | \s)$ will include an implicit intercept for each connected component of the graph in Bayesian analysis. For simplicity, this paper will assume that the $G$ is connected and therefore there is a single global intercept included in the ICAR prior.

 \citet{Reich} investigated in depth the effect multicollinearity between $\X$ and the spatial random effect induced by the ICAR prior has on $E(\betas | \Y)$ and $\textrm{Var} (\betas | \Y)$ using a re-parameterization of the ICAR model: 
 \begin{align} \label{ICAR}
\bm{Y} = \bm{X} \betas+ \bm{V} \bm{\delta} + \bm{\epsilon}, \\ 
 p( \bm{\delta} | \tau_s )  \propto \tau_s^{\kappa} \exp \left\{ \frac{- \tau_s}{2} \bm{\delta}^T \bm{\Lambda} \bm{\delta} \right\} \nonumber,
 \end{align}
where  $ \bm{V} \bm{\Lambda} \bm{V}^T$ is the eigendecomposition of $\bm{Q}$, and $\bm{\Lambda} = \textrm{diag}(\lambda_1, \ldots, \lambda_n)$ where $\lambda_1 \geq \lambda_2, \ldots \geq \lambda_n$. Under the assumptions of this paper (i.e., that $G$ is connected), $\kappa = \frac{n-1}{2}$.
 Using a Bayesian approach with flat priors on $\bm{\betas}$ and independent gamma priors on $\tau_s, ~ \s$, \citet{Reich} and \citet{Hodges2} illustrated that $E [ \bm{\betas} |Y, \s, \e] \neq E[\betass{NS} | \Y ]$ in the presence of multicollinearity between $\X$ and $\bm{V}$.  When regression coefficients are of primary interest, they argued that spatial random effects are added merely to account for spatial correlation in residuals when computing the posterior variance. Under this reasoning, the spatial random effect should not change the point estimates of the fixed effects. Furthermore, \citet{Reich} suggested that this multicollinearity causes ``overinflation" of the posterior variance of the regression coefficients. This overinflation was noted to be exacerbated when $\X$ was correlated with low frequency eigenvectors of the graph Laplacian. In spatial regression, \citet{Reich} argued that two factors can lead to an increase in the posterior variance of regression coefficients as compared to the NS model: 1) collinearity with the spatial random effects and 2) a reduction in the effective number of observations because of spatial clustering. They proposed a model that they claimed would eliminate the former of these factors while preserving the latter. In other words, the RHZ model was designed to alleviate spatial confounding according to \cref{def:spatconf}:

\begin{enumerate}
	\item $E(\betass{RHZ} | \Y) = E(\betass{NS} | \Y)$
	\item $ \textrm{Var}({\betass{NS,}}_i | \Y) \leq  \textrm{Var}({\betass{RHZ,}}_i | \Y) \leq  \textrm{Var}({\betass{ICAR,}}_i | \Y) $ for $i= 1, \ldots, p$.
\end{enumerate}
 
The authors provided theoretical support for first inequality of the second claim.\footnote{Section 3 of \citet{Reich} provides a mathematical argument that this should be true. However, the argument appears to rely on the stochastic ordering of a gamma distribution. It appears the authors assume they are using a gamma distribution with a scale parameter- which is stochastically increasing in the scale. Their work suggests they are instead working with a rate parameterization- which is stochastically decreasing in the rate parameter.}  To accomplish these goals, the authors proposed using synthetic predictors which are orthogonal to the column space of $\X$. Define $\Px = \X ({\X}^T \X)^{-1} \X^T$ to be the projection matrix onto the column space of $\X$, and $\Pperpx = \bm{I} - \Px$ to be its orthogonal complement. Let $\bm{L}$ be the $n \times (n-p)$ matrix composed of the eigenvectors of $\Pperpx$ associated with an eigenvalue of $1$ and let $\bm{K}$ be the $n \times p$ matrix composed of the eigenvectors associated with an eigenvalues of $0$. Model  \eqref{ICAR} can be re-written as a function of $\bm{\delta}_1=  \bm{L}^T \bm{V} \bm{\delta}$ and $\bm{\delta}_2= \bm{K}^T \bm{V} \bm{\delta}$. The RSR method proposed by \citet{Reich} involves setting $\bm{\delta}_2 =\bm{0}$. Using this notation, the RHZ model can be expressed as:
 \begin{eqnarray} \label{RHZ}
\bm{Y} = \bm{X} \betas + \bm{L} \bm{\delta}_1 + \bm{\epsilon},  \\ 
 p( \bm{\delta}_1 | \tau_s )  \propto \tau_s^{\kappa} \exp \{ \frac{- \tau_s}{2} \bm{\delta}_1^T \bm{Q_s} \bm{\delta}_1 \} \nonumber,
 \end{eqnarray}
 where $\bm{Q_s} = \bm{L}^T \bm{Q} \bm{L}$. \citet{Hodges2} explained that this solution assigns all variability explained by both $\bm{V}$ and $\bm{X}$ to $\bm{X}$ by restricting spatial smoothing to the orthogonal complement of the fixed effects.
 
 \citet{Hughes} also argued that variance inflation caused by spatial confounding is a concern in the SGLMM. However, they stated that the RHZ model \eqref{RHZ} fails to account for the underlying graph in the construction of $\bm{L}$, thereby permitting structure in the spatial random effects that corresponds to negative spatial dependence. \citet{Hughes} suggested such negative spatial dependence should not be expected in the context in which spatial models are generally fit. They developed a model (hereafter the HH model) that they stated would eliminate negative dependence while alleviating the impact of spatial confounding on the marginal posterior variances. Utilizing a Bayesian framework with proper priors on the regression coefficients, they suggested that rather than using  $\bm{L}$, one should instead use $\bm{M}_q$, where $\bm{M}_q $  is the $n \times q$ matrix composed of the eigenvectors of what they referred to as the Moran operator $\Pperpx \bm{A}  \Pperpx$. In support of the argument that $\bm{M}_q$ should be preferred to $\bm{L}$, the authors pointed out that the column vectors of $\bm{M}_q$ had more spatial structure than the column vectors of $\bm{L}$. 
 
 \citet{Hughes} also stated that this model naturally lent itself to dimension reduction. The Moran operator is the numerator of a generalization of the Moran's $I$ statistic, a popular non-parametric measure of spatial dependence introduced by \citet{moran1950notes}. We will refer to this generalization of the Moran's $I$ as $I_{\X}(\bm{A})$. Relying on work on the original Moran's $I$ statistic in the context of bounded regular tessellations by \citet{Boots}, \citet{Hughes} noted the (standardized) spectrum of the Moran operator comprises all possible values of $I_{\X}(\bm{A})$ and its eigenvectors comprise of all possible mutually distinct patterns of clustering after accounting for $\X$ and the graph. Importantly, they argued that choosing only ``attractive" eigenvectors (those associated with positive eigenvalues) of the Moran operator would improve the inference on regression coefficients by eliminating patterns of ``repulsive" spatial dependence (eigenvectors associated with negative eigenvalues). Furthermore, they suggested that for most graphs only selecting the first $.1 n$ eigenvectors should be sufficient to perform well for regression. The work in \citet{Hughes} relied on simulations to illustrate the points made. 
 
 A downside to both the RHS and HH models is a loss of some of the computational efficiency inherent in the ICAR prior utilized in \eqref{ICAR}. ICAR models capture spatial dependence through a Gaussian Markov Random Field (GMRF). This is appealing because the precision matrix for the random field is very sparse, which allows the use of sparse matrix routines to facilitate efficient computations \citep{Rue,Pac}. The matrix $\bm{Q}_s$ need not be, and generally is not, sparse.  Furthermore, the RHZ and HH methods do not extend naturally to spatial models that do not utilize the ICAR prior.  Recent work by \cite{Prates} attempts to to overcome the computational limitations of RSR methods. Their work suggests projecting the vertices of the graph $G$ onto the orthogonal space of the design matrix $\X$. Using the projected vertices, they then construct a new sparse precision matrix $\bm{Q}^{\perp}$ for use in analysis. This model is referenced as the PAR model in \cref{tab1}. \citet{guan} also suggested a projection based approach to approximate covariance matrices of spatial random effects to improve the computational efficiency of RSR methods. This method can also be extended to spatial models beyond the ICAR model. 

 RSR methods are not universally accepted; some find the assumption that the spatial random effect operates orthogonally to the fixed effects to be too strong \citep[See e.g.,][]{Pactwo}. Furthermore, recent work has highlighted that RSR methods can suffer from elevated rates of Type-S errors under model misspecification \citep{Hanks,Prates}. Type-S error is the Bayesian analogue to Type I error, and we define it to occur when the equal-tailed 95\% credible interval for a regression coefficient that is truly 0 does not include 0. The model misspecification studied has involved a generating model which include spatial random effects that do not operate orthogonally to the fixed effects. Currently, the prevalent belief is that the elevated Type-S error is caused by such misspecification. However, in this paper we will show that the phenomena occurs even without it.

 It remains an open question when RSR methods should be preferred over traditional spatial models. Understanding when RSR methods perform well is important because RSR methods can lead to a very different interpretations of the effect of regression coefficients \citep{Hanks}. Attempts to understand when RSR methods are appropriate have led to suggestions that when $\X$ is correlated with low frequency eigenvectors of the graph Laplacian, RSR methods  should be preferred (\citet{Reich,Prates}, \citet{Hefley}).
 
\section{Consequences of Orthogonal Smoothing} \label{mywork}

In this section, we investigate the impact that methods designed to alleviate spatial confounding have on inference for regression coefficients.  Formally, we state the following definition of a RSR model:
\begin{defn}
	A \textit{Restricted Spatial Regression Model} is any model of the form \eqref{second} with $\W$ chosen such that  $\mathcal{C}(\W) \perp \mathcal{C}(\X)$. 
\end{defn}
Here and throughout the paper $\mathcal{C}(\cdot)$ denotes the column space of a matrix. Both the RHZ and HH models are special cases of this class of models. Note that the ICAR model is not a special case of a RSR model. For RSR models, any results stated for $\betas$ (relying on $\X$) are also true for $\bm{\beta}^*$ (by replacing $\X$ with $\X^*$). Since the PAR model is an attempt to approximate these methods, we will not spend time directly investigating its performance. All of the following results assume a Bayesian analysis that uses flat priors on the regression coefficients and independent gamma priors on the precision parameters $\s$ and $\e$ with respective shape parameters $\at, ~ \as > 0$ and scale parameters $\bt, ~ \bs > 0$. 

\subsection{Properties of Marginal Posterior Distribution}

To begin to investigate the impact RSR models have on inference for regression coefficients, we consider the mean and variance of the marginal posterior distribution of $\betas$. As previously noted, RSR models are developed in the areal data setting with the expectation that the marginal posterior mean of $\betas$ will be $({\X}^T \X)^{-1} \X^T \Y$. Working in a geostatistical setting and utilizing an empirical Bayes approach, \citet{Hanks} observed this would be the case for a RSR method in that context as well. However, all of the previous work has relied either on observations regarding distributions of $\betas$ conditional on the precision parameters (for example, \citet{Reich,Hodges2}) or on the full conditional distribution of $\betas$ (\citet{Hanks}). \cref{thm:Thm1} offers the first rigorous proof of the existence of the mean and variance of the marginal posterior distributions of $\betas$ in RSR models, as well as the forms these moments take. It illustrates that the point estimates obtained from RSR models will be the same as the non-spatial point estimates. It also shows the variance of the distribution looks functionally similar to the variance we would have obtained in the non-spatial model. For the rest of the paper, we adopt the following notational shortcut: $ \sigma_{\epsilon}= \frac{1}{\tau_{\epsilon} } $.

\begin{thm} \label{thm:Thm1}
	Under conditions A.1[1]- A.1[5], a RSR model with $\W$ a $n \times q$ matrix will give rise to the marginal posterior distributions of $\betas$ such that: 
	\begin{eqnarray*}
		\textrm{E} [\bm{\betas} | \Y ] &=& ({\X}^T \X)^{-1} \X^T \Y \\
		\textrm{Var} [ \bm{\betas} | \Y ]  &=& ({\X}^T \X)^{-1} E[\frac{1}{{\e}^{RSR}} | \Y]  = ({\X}^T \X)^{-1} E[{\sigma}_{\epsilon}^{RSR} | \Y].
	\end{eqnarray*}
	\noindent \textit{Proof}:  See Appendix Section B.1.
\end{thm}

An immediate consequence of \cref{thm:Thm1} is that the dimension of $\bm{W}$ will not impact the point estimates for the regression coefficients. Thus, any choice of $q$ linearly independent column vectors which are orthogonal to $\mathcal{C}(\X)$  will give the same point estimates as a HH model using $\bm{M}_q$. Of course, incorporating spatial dependence in a model is a computationally expensive and often theoretically complicated endeavor. If the concern were simply to obtain point estimates from a non-spatial model, there would be no need to use a spatial model.  As noted previously, the focus in fitting a spatial model is often to account for spatial correlation in residuals when computing the posterior variance. Hence, our interest lies in $\textrm{Var} [ \bm{\betas} | \Y ]$, which \cref{thm:Thm1} illustrates is a function of $E[{\sigma}_{\epsilon}^{RSR} | \Y] $.

The prevalent expectation is that the marginal posterior variances of the regression coefficients in RHZ and HH models will be greater than the marginal posterior variances obtained in the non-spatial model. In settings where RSR models are currently recommended for use, \cref{thm:Thm2} illustrates that for a broader class of RSR models including the RHZ and HH model, the opposite is true. In the statement of \cref{thm:Thm2}, $r = \frac{\s}{\e}$ is the inverse of the signal-to-noise ratio.

\begin{thm} \label{thm:Thm2}
	Under conditions A.1[1]-A.1[4], a RSR model with $\W$ a $n \times q$ matrix with orthonormal columns and $\F$ a positive definite and symmetric matrix will always result in a marginal posterior variance for ${\betas}_i, \, i = 1, \ldots, p$ that is less than or equal to that of the posterior variance that would have been obtained in the non-spatial model whenever the following holds:\\
	\[ \frac{\textrm{E}[ r | \Y]}{E[ {\s} ]} < E \left[ \frac{1}{ {\e}^{NS}} | \Y  \right] = E \left[ \sigma_{\epsilon}^{NS} | \Y \right], \]
	where $E[ {\s} ]$ is the prior expectation of $\s$ under the RSR model. 
	%i.e., $\textrm{E} (\sigma^{W} | \Y ) \leq \textrm{E} (\sigma^{NS} | \Y )$
		
	\noindent \textit{Proof}:  See Appendix Section B.2.
\end{thm}

As a practical matter, the conditions set forth in \cref{thm:Thm2} will often be met. To see this, consider that spatial regression models, including RSR models, are typically used in settings in which the $\X \betas$ in the NS model does not explain nearly all the variability in $\Y$ (i.e., when  $E \left[ \sigma_{\epsilon}^{NS} | \Y \right]$ is not nearly zero). Spatial confounding is thought to be problematic in settings where $r$  is ``small" for traditional spatial regression models (see e.g., \citealt{Reich} and \citealt{Hodges2}). ``Small" is usually taken to mean substantially less than $1$ (see e.g., \citealt{Reich} and \citealt{Prates} who consider values of $r= .1$ and $.2$ respectively). This corresponds to situations where there is residual spatial dependence unexplained by $\X$ (and hence larger $E \left[ \sigma_{\epsilon}^{NS} | \Y \right]$ ). The better a RSR model is at explaining this residual spatial dependence, the smaller $\textrm{E}(r| \Y)$ will be.  Finally, in the literature, $E[ {\s} ]$ ranges in value from $1$ \citep[see e.g.,][]{Hodges2} to $1,000$ \citep[see e.g.,][]{Hughes}. Thus, the conditions in \cref{thm:Thm2} will typically be satisfied for prominent examples of RSR models.  

\cref{thm:Thm2} offers the first fully rigorous proof detailing the bounds on the variance of the marginal posterior distribution of $\betas$ for RSR models. Working in a geostatistical setting with an empirical Bayesian analysis, \citet{Hanks} stated without proof that the inequality related to posterior variances in \cref{thm:Thm2} should hold for a particular RSR model. However, perhaps because of the difference in settings, this observation has not been perceived as contradicting the expectations for the more commonly studied RHZ and HH models, which are precisely the opposite. Interestingly, the work in \citet{Hanks} is often reported as illustrating that RSR methods can have elevated Type-S error rates \textit{under model misspecification} \citep{Prates,page2017estimation}. \citet{Prates} further noted that these elevated Type-S error rates were exacerbated in settings in which $r$ is small. Yet, \cref{thm:Thm2} suggests that RSR models will have elevated Type-S error rates in the absence of model misspecification. 
In a frequentist setting, the OLS estimators would be normally distributed and an ordering of variance with the same mean would suggest that the $100 (1- \alpha)\%$ confidence intervals would be nested (with the interval of the distribution with less variance completely contained in the other). In that setting, \cref{thm:Thm2} would indicate that any RSR model would have higher rates of Type I error and lower rates of Type II error than the non-spatial model. Since the non-spatial model yields estimators with sampling distributions whose coverage probability matches the nominal coverage probability, this indicates that the RSR coverage probability would always be less than the nominal coverage probability under the conditions of \cref{thm:Thm2}. Hence, \cref{thm:Thm2} would explain the elevated levels of Type-S error observed for RSR models, and it would mean these elevated levels could exist even without model misspecification.

In Bayesian analysis, when inference on regression coefficients is of primary interest, practitioners typically utilize an equal-tail credible interval. This is especially true in settings like those of this paper where improper priors preclude the use of Bayes factors. Because the marginal posterior distribution $f(\betas | \Y)$ is not normal for any of the models discussed, the variance ordering in  \cref{thm:Thm2} need not necessarily indicate a relationship between the credible intervals.

Investigating the relationship between equal-tailed credible intervals for the non-spatial model and a RSR model for all possible choices of graph, data, and spatial basis vectors is difficult because these distributions will not generally be available in closed form. However, for the case that $\beta^X \in \mathbb{R}$, it is possible to make some observations that shed light on the relationship between the credible intervals of a RSR model and the NS model. This is formalized in \cref{thm:Thm3}.

\begin{thm} \label{thm:Thm3}
	Assume $\beta^X \in \mathbb{R}$, A.1[3] holds, and A.1[5] holds. Let $g(\beta^X | \Y)$ be the marginal posterior probability distribution (pdf) from a RSR model with choice of $\F$ which is a symmetric and positive definite $q \times q$ matrix. Let $h(\beta^X | \Y)$ be the marginal posterior pdf from the non-spatial model. Then, $ g(\beta^X | \Y) = \textrm{O}(h(\beta^X | \Y))$ as $\betax \to \infty$ and $\betax \to - \infty$.\\
		\noindent \textit{Proof}:  See Appendix Section B.3.
\end{thm}

An immediate corollary to this result is the following.

\begin{cor} \label{cor:cor1}
	Define $G$ and $H$ to be the respective cumulative distribution functions of $g(\beta^X | \Y)$ and $h(\beta^X | \Y)$.  $\exists ~ C > 0$ such that:
	
	\begin{enumerate}
		\item $\limsup\limits_{\beta^X \to - \infty} \frac{G(\beta^X | \Y)}{H(\beta^X | \Y)} \leq C, $
		\item $\limsup\limits_{\beta ^X\to \infty} \frac{1- G(\beta^X | \Y)}{1- H(\beta^X | \Y)} \leq C $
	\end{enumerate}
	
	Furthermore, for $D_g^*$ and $K(C_q,D_h)$ as defined in the proof in \cref{proofs}, if $D_g^* > K(C_q,D_h)$, then $\exists ~  \beta^* > 0, ~ \beta_* < 0$ such that:
	
	\begin{enumerate}
		\item $G(\beta | \Y)   \leq H(\beta | \Y) ~ \forall \beta < \beta_*$
		\item $ 1- G(\beta | \Y) \leq 1 - H(\beta | \Y) ~ \forall \beta > \beta^* $
	\end{enumerate}
	\noindent \textit{Proof}:  See Appendix Section B.3.	
\end{cor}

Generally, what these results indicate is that the tails of the marginal posterior distribution for the non-spatial model and the tails of the marginal posterior distribution for RSR models decay roughly at the same rate. The combination of \cref{thm:Thm1} and \cref{thm:Thm3} suggests that that RSR models offer inference very similar to a non-spatial model with respect to regression coefficients. Importantly, these results are invariant to the choice of graph, data, and spatial basis vectors.  To see how the inference will be similar, note the posterior distribution $f(\beta^X_{NS} | \Y)$ is symmetric about its mean. While, $f(\beta^X_{RSR} | \Y)$ is not symmetric about its mean, two applications of Jensen's inequality can bound the distance between a mean and median by the standard deviation. In practice, the standard deviation for a RSR model is always quite small. Thus, RSR and NS models produce marginal posterior distributions with a similar medians and similar tail decay. This result is important because RSR models are not offering much beyond what an NS model offers, yet they are much more computationally expensive. For example, \citet{Hughes} reported that a NS model took under a minute, while their dimension reductions technique took over 6 hours. 

If the condition put forth in Corollary \eqref{cor:cor1} holds, it also means that for sufficiently large credible intervals, the credible interval for the RSR model will be completely contained in the corresponding credible interval for the non-spatial model.  Importantly, these results require no assumptions about $\textrm{E}(r | \Y)$. Just as in the frequentist setting, this means that RSR methods will only capture the regression coefficient if the non-spatial model does. Furthermore, the RSR methods will always have higher rates of Type-S error than the non-spatial model - even if there is no spatial confounding. However, in practice determining whether the condition in Corollary \eqref{cor:cor1} is satisfied will be impractical. Therefore, we further the investigation with simulation studies in \cref{simu}.

\subsection{Implications for Inference} \label{sec32} The premise of RSR  models is that incorporating spatial dependence will yield better inference on the regression coefficients. In particular, the HH model and those influenced by it have emphasized choosing spatial basis vectors which are spatially smooth. This emphasis is inherited from work on dimension reduction techniques developed for spatial process modelling where this technique is quite useful \citep[See e.g.,][pages 353-354]{cressie2015statistics}. 
Recall our definition of RSR models includes both models designed to capture spatial dependence as well as models with arbitrary choices of $\W$ and $\F$.
Yet, the results of \cref{thm:Thm1} -- \cref{thm:Thm3} hold whether the models are designed to incorporate spatial dependence or not. A natural question then is: For RSR models, do choices of $\W$ (and perhaps $\F$) which exhibit spatial patterns behave any differently than those which do not? 

\cref{thm:Thm4} begins to explore the answer to this question. Recall that one of the criticisms of the RHZ model \eqref{RHZ} is that the basis vectors $\bm{L}$ do not appear to be spatially smooth. 
\cref{thm:Thm4} states that when $\bm{F}$ is of the form $\bm{W}^T \bm{Q} \bm{W}$, the inference on the regression coefficients will be invariant to $\mathcal{C}(\bm{W})$. This result suggests that the choice of basis vectors which exhibit spatial patterns (such as those from the Moran operator) does not affect inference on regression coefficients in the ways expected. As an example, for moderate $q$, given a selection of spatially smooth basis vectors, we could pick a different basis of the column space (which may not appear spatially smooth), and inference for $\betax$ would be equivalent.

\begin{thm} \label{thm:Thm4}
	A RSR model with ${\W}_1$ a $n \times q$ matrix with orthonormal columns and ${\F}_1 =  {\W}_1^T \bm{B} {\W}_1$ for arbitrary non-null symmetric $\bm{B}$ will yield the same marginal posterior distribution  $f(\betas |\Y)$ as any other choice  ${\W}_2$ with orthonormal columns such that $\mathcal{C}({\W}_1) = \mathcal{C}({\W}_2)$ and ${\F}_2 = {\W}_2^T \bm{B} {\W}_2$.\\
		\noindent \textit{Proof}:  See Appendix Section B.4.
\end{thm}

\cref{thm:Thm4} suggests that the assumptions underlying the development of RSR models may be worth re-evaluating. 
The addition of the spatial basis vectors in RSR models is assumed to make the regression coefficient estimates less biased and more precise. 
However, further investigation into RSR models indicates that by choosing these spatial basis vectors to be orthogonal to the column space of $\X$, RSR models are effectively transforming a mixed-effects model into an overfit fixed effects model. Thus, RSR models may not be aiding in inference for regression coefficients as expected.

To understand the intuition behind this last claim, it is instructive to take a moment to review the non-spatial Gaussian model. The method of ordinary least squares makes the assumption that all the variability in the direction of $\X$ can be described by the linear combination of $\X \betas$. This assumption results in fitted values which are simply a projection of $\Y$ onto the column space of $\X$. In other words, the point estimates are $\hat{\betas} = ({\X}^T {\X})^{-1} {\X}^T \Y$. Importantly, this assumption is also reflected in the estimated variance of these point estimators. This can be seen in the familiar form for the estimated variance of the OLS estimators, where the  $i^{th}$ regression coefficient $\hat{{\betas}}_i$ is simply the $(i,i)$ element of $({\X}^T {\X})^{-1} \hat{\sigma}_{\epsilon}^{NS}$:
\[   \widehat{ \textrm{Var}} (\hat{\betas}_i) = ({\X}^T {\X})_{ii}^{-1} \hat{\sigma}_{\epsilon}^{NS} = ({\X}^T {\X})_{ii}^{-1} \frac{ || \Pperpx \Y ||^2 } {n-p}.   \]
The first component reflects the relationship between the different covariates, which is not of interest in this paper. The second term, however, is a consequence of the assumption that the variation in the direction of $\X$ can be explained by $\X \betas $. Here, $\hat{\sigma}_{\epsilon}^{NS}$ is simply a function of the magnitude of the component of $\Y$ unexplained by any linear combination of the column space of $\X$.

In Bayesian analysis, the assumption that all the variability in the direction of $\X$ can be explained by a linear combination of the columns of $\X$ has analogous implications. For the non-spatial model, the point estimates for the regression coefficients will simply be those of the OLS model. The associated posterior variance for the $i^{th}$ regression coefficient ${\betas}_i$  will be the $(i,i)$ element of $ ({\X}^T {\X})^{-1} \textrm{E} (\sigma_{\epsilon}^{NS} | Y) $:
%$\textrm{E} [ \betass{NS}| \Y]  = ({\X}^T {\X})^{-1} {\X}^T \Y$:
\begin{equation} \label{BayesianNS}
\textrm{Var} [ \betass{NS,i}| \Y]  = ({\X}^T {\X})_{ii}^{-1} \textrm{E} (\sigma_{\epsilon}^{NS} | Y) = ({\X}^T {\X})_{i,i}^{-1} \frac{ \bs^{-1} + .5 || \Pperpx \Y||^2}{ (\as - 1) + .5 (n-p)  }.   
\end{equation}
Although a bit messier than the OLS setting, this relationship again shows that the uncertainty associated with the regression coefficients is a function of the magnitude of the component of $\Y$ unexplained by the the column space of $\X$. For $ n \gg p$ and the typical choices of hyper-parameters ($\as, \bs^{-1} < 1$), this will in fact be quite similar to the estimate obtained via OLS.

In a fixed effects model, researchers are well aware that overfitting the model with the addition of synthetic predictors can distort the true relationship between $\Y$ and $\X$. For example, consider the non-spatial model with new design matrix $\bm{T} = [ \X ~ \W]$, with $\mathcal{C}(\X) \perp \mathcal{C}(\W)$. Because $\W$ is constructed to belong to the orthogonal complement of the columns space of $\X$, the point estimates remain unchanged. If $\W$ was a $n \times (n - p)$ matrix of rank $n-p$, $\bm{T}$ would be a basis of $\mathbb{R}^n$. Thus, by construction, $\Pperpt \Y = \bm{0}$. More generally, because $({\bm{T}}^T {\bm{T}})^{-1}$  is a block diagonal matrix with diagonals $({\X}^T {\X})^{-1}$ and  $({\W}^T {\W})^{-1}$, the following will be true:
\begin{equation*} \label{NSwW}  \widehat{ \textrm{Var}} (\hat{\betas}_i) = ({\X}^T {\X})_{ii}^{-1} \hat{\sigma_{\bm{T}}} = ({\X}^T {\X})_{ii}^{-1} \frac{ || \Pperpt \Y ||^2 } {n-p-q}.
\end{equation*}
For fixed $\X$,  $|| \Pperpt \Y ||^2$ is monotonically decreasing as a function of the number of columns of $\W$. If we add enough (linearly independent) columns to $\W$, we will eventually explain the variation in $\Y$. Of course, the $n-p-q$ in the denominator ensures that $\hat{\sigma_{\bm{T}}}$ is not monotonically decreasing as the number of columns of $\W$ increases. However, a basis for the orthogonal complement to $\mathcal{C}(\X)$ can still be added in such a way as to ensure that $\hat{\sigma_{\bm{T}}}$ will be very, very close to zero (see \cref{overfitexample} for an example).  Thus, even if there is truly no relationship between $\Y$ and $\X$, with enough synthetic covariates this approach can essentially ensure that the covariates are deemed ``significant" (assuming that the columns $\X$ are not very strongly correlated). Of course, as noted this problem is well-understood in fixed effect models, and this approach is not considered acceptable in the practice of regression analysis. 

Returning to RSR models, \cref{thm:Thm1} illustrates that like the NS model, the point estimates for these models assume that all the variation in the direction of $\X$ can be explained by the linear combination $\X \betas$. However, \cref{thm:Thm2} suggests that when it comes to the posterior variance, RSR models are going beyond this assumption. In RSR models, $\F$ and $\s$ act as regularization parameters that can smooth some of the elements of $\bm{\delta}$ to 0. These terms make it difficult to find a closed form expression for $E[ \sigma_{\epsilon}^{RSR} | \Y]$. However, we can still gain intuition by looking at the behavior of $E[ \sigma_{\epsilon}^{RSR} | \Y, r]$, where $r=\frac{\s}{\e}$ is again the inverse of the signal-to-noise ratio. Straightforward calculations show that $E[ \sigma_{\epsilon}^{RSR} | \Y, r]$ can be expressed as: 
\[ \frac{  \frac{1}{\bs} + \frac{r}{\bt} + .5 {\Y}^T  (\Pperpx - \W (\bm{I}  + r  \bm{F}  )^{-1}{\W}^T ) \Y }{  \as + \at +  .5 (n-p)-1}. \]
Note that the $(n-p)$ term in the denominator will remain unchanged no matter how many columns are added to the spatial basis matrix $\W$. If we consider the behavior of $E[ \sigma_{\epsilon}^{RSR} | \Y, r]$ as the signal-to-noise ratio increases (i.e., $r \to 0$) for $\W$ a $n \times (n-p)$ matrix of full rank, then 
\[ \lim_{r \to 0} E[ \sigma_{\epsilon}^{RSR} | \Y, r] = \frac{  1}{ \bs (\as + \at+ .5 (n-p)-1)}.  \]
For moderate $p$ (relative to $n$) and the typical choices of $\as$ and $\bs$, this value will be quite small. Note that this is slightly smaller than the value that would be obtained in a Bayesian NS model if the fixed effects $\X$ alone completely explained the variation in $\Y$ (see \cref{BayesianNS} for the case when $|| \Pperpx \Y||^2 = 0$). However, this reduction is a function of how well the additional columns of $\W$ explain the variability in $\Y$, and the signal-to-noise ratio will tend to increase as ``useful" columns are added to $\W$. In other words, if $\X$ did a poor job of explaining the variability of $\Y$, the addition of columns to $\W$ will lead to a more dramatic decrease in $E[ \sigma_{\epsilon}^{RSR} | \Y]$ and hence $\textrm{Var}[\betas | \Y]$. Because of the regularization induced by $\F$ and $\s$, this decrease will not necessarily be monotonic. 

To understand the implications of this discussion as well as \cref{thm:Thm4} on inference, we analyze a real world dataset with a series of RSR models. We consider the SAT data originally analyzed by \citet{wall2004close}. We use the data set provided as an example in \citet{bivand2008applied}. The data include statewide averaged verbal scores on the SAT and the percent of students eligible to take the test in 1999 for the 48 contiguous states of the United States. Let $y_i$ denote the statewide averaged SAT verbal score for the $i$th state and $\bm{Y} = (y_i, \ldots, y_{48})$. If $\bm{X}_1$ = Percent of students eligible to take the exam, then we consider a model with covariates $\X^* = [\bm{1} ~ \bm{X}_1 ~ \bm{X}_1^2]$ with associated regression coefficients $\beta_0, ~ \beta_1, ~ \beta_2$. We fit a series of HH models for various $M_q, ~ q= 1, \ldots, 45$. For all these analyses, $\at= .5, ~ \bt=2000$ as suggested by \citet{Hughes}. 

\begin{figure}[H] 
	\centering
	\subcaptionbox{}{\includegraphics[width=0.30\textwidth,height=.22\textheight]{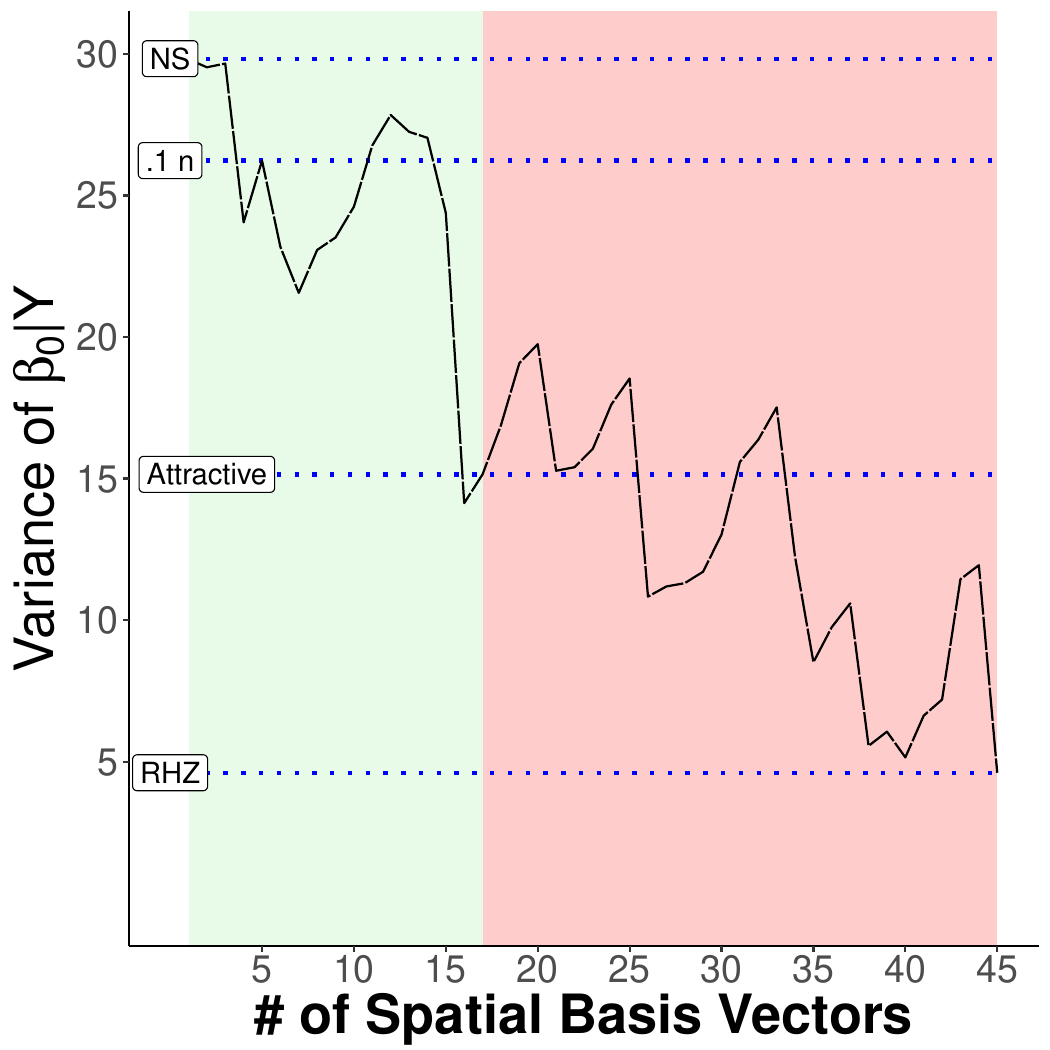}}%
	\hfill % <-- Seperation
	\subcaptionbox{}{\includegraphics[width=0.30\textwidth,height=.22\textheight]{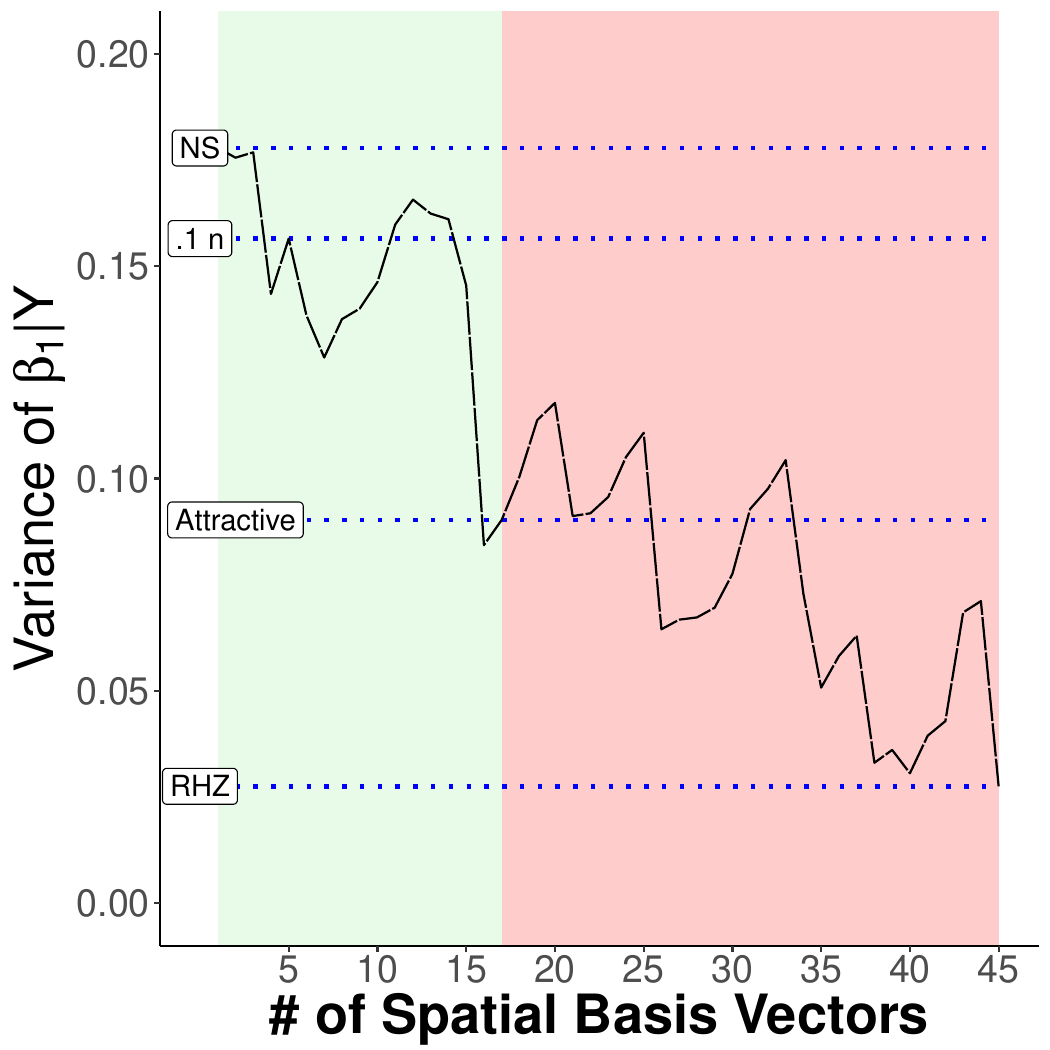}}%
	\hfill % <-- Seperation
	\subcaptionbox{}{\includegraphics[width=0.30\textwidth,height=.22\textheight]{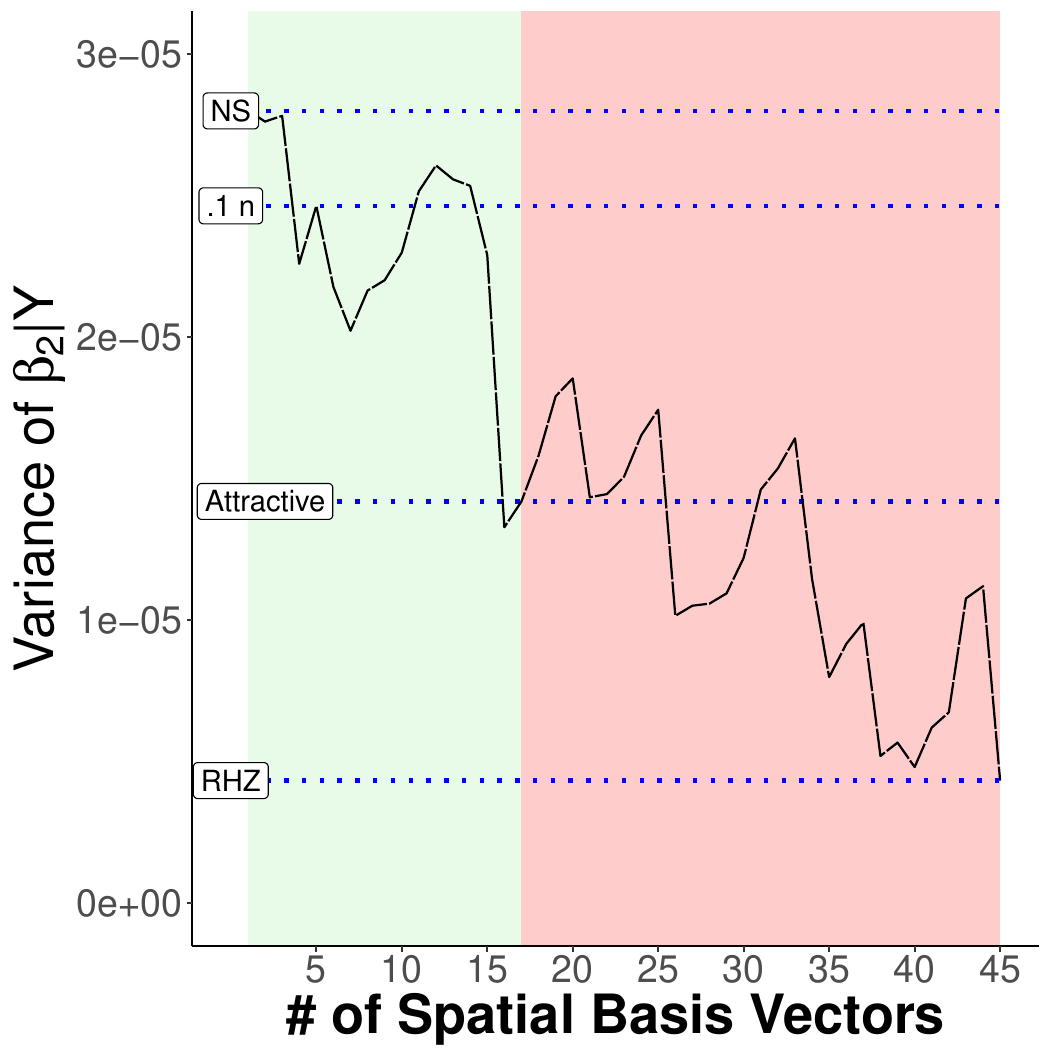}}%
	\caption[Description] 
	{  These graphs depict the posterior variance of each of the regression coefficients. Green indicates spatial basis vectors which are ``attractive." Red indicates spatial basis vectors which are ``repulsive."  These figures were constructed using the ggplot2 R package \citep{ggplot2}. }
	\label{fig:maps}
\end{figure}

In \cref{fig:maps}, we plot the marginal posterior variance for $\beta_0$,~ $\beta_1$, and $\beta_2$ for each of the various HH models.  The green in the graphs represents vectors which are ``attractive" and the red represents vectors which are ``repulsive." All RSR models result in approximately the same point estimates as those obtained as in the non-spatial model $(590.5, -2.84, .022)$ and have  marginal posterior variances less than those obtained in the non-spatial model. Based on the discussion above, we expect that as the number of columns of $\W$ increases, the posterior variances should tend to decrease. And in fact, that seems to be the general pattern. Since the ``attractive" vectors are added to the model first, these tend to have a larger variance than models that include both ``attractive" vectors and ``repulsive" vectors. Note that this same pattern can be obtained by adding arbitrary synthetic covariates to a non-spatial model (see, \cref{overfitexample}).

However, note, that the distinction between ``attractive" and ``repulsive" eigenvectors is not directly linked to the posterior variances otherwise.  For example, the posterior variance obtained for the set of ``attractive" vectors proposed by \citet{Hughes} is achieved again for choices that include up to $14$ ``repulsive" vectors, but is not obtained with the recommended dimension reduction to $.1 n$. A restriction to ``attractive" eigenvectors was originally proposed as a way to approximate the inference from the RHZ model \citep[See, Section 7 of][]{Hughes}. However, there is no clear association between the variances that are obtained via the RHZ model and those obtained restricting the spatial basis vectors to ``attractive" ones.

It has been claimed that RSR models assume that all the variability in the direction of $\X$ can be described by the linear combination $\X \betas$ \citep{Hodges2,Hanks}. In this section, we point out that the NS already makes this assumption.  In order to gain an intuitive understanding of why the posterior variances are RHZ models are always bounded above by the posterior variance of the corresponding NS model, we offer an alternative interpretation which indicates the the reduction in the posterior variance for RHZ models is akin to what would be observed an overfit fixed effects model. When this intuition holds, we would expect that RHZ models would always have a higher Type-S error rate than NS models even in the absence of model misspecification. Importantly, these problems cannot be resolved by selecting different spatial basis vectors within the class of RSR models, and dimension reduction techniques designed to mimic smoothing orthogonally to the fixed effects will likely suffer from the same problems. However, just as is the case for the implications of \cref{thm:Thm3}, exploring this rigorously is impractical. Therefore, in the next section we use simulations to further our investigation.

\section{Simulation Studies} \label{simu}
In this set of simulations, we investigate the relative performances of the ICAR model, the RHZ model, and the NS model for continuous areal data and count areal data. In particular, we explore how and when the RHZ model offers inference different than the inference which would have been obtained with a NS model. We also explore whether RHZ models improve the inference for regression coefficients relative to the ICAR model in the presence of spatial confounding.

Currently, spatial confounding is thought to be a concern when $\X$ is spatially dependent and there is residual spatial dependence unexplained by the covariates. Therefore, all of the following simulations involve covariates which are spatially dependent and residual spatial dependence (either with a spatial random effect or with ``unobserved" covariate). To simulate the covariates, we rely on previous work in spatial confounding. In the context of areal data, statistics developed by \citet{Reich} and \citet{Prates} essentially define spatial dependence in $\X$ to be the existence of correlation between $\X$ and low frequency eigenvectors of the graph Laplacian. These statistics reflect a common assumption in the literature that the eigenvectors of the graph Laplacian $\bm{Q}$ associated with low eigenvalues exhibit spatially smooth patterns while those associated with high eigenvalues oscillate rapidly. In the continuous case, this phenomenon can be formalized. For the low frequency eigenfunctions,  Bernstein estimates show smoothness (see e.g., \cite{zelbook} Theorem 5.17); whereas for  high frequency eigenfunctions, rapid oscillation can be shown using upper bounds on the size of nodal domains (see e.g., \cite{zelbook} Theorem 13.1). We are not aware of any formalization of the phenomenon in the discrete case. However it does seem true in practice, and we will utilize this assumption in the following work. Therefore, in the following simulations we generate covariates that are correlated with low frequency eigenvectors of the graph Laplacian. To do so, we utilize the fact that on a connected graph, any column vector $\bm{l} \in \mathbb{R}^n$ can be written as follows:
\begin{eqnarray}
s_l \sqrt{n-1}  \bm{V} \bm{\rho}_{l V} + \bm{1} \bar{l}  \label{decomp}
\end{eqnarray}
where $s_l$ and $\bar{l}$ are the sample standard deviation and sample mean of $\bm{l}$, $\bm{\rho}_{l V} =  (\rho_{l, V_i}, \ldots ,\rho_{l,V_{n-1}}, 0)^T$ is a column vector where $\rho_{l,V_{i}}$ is the correlation between $\bm{l}$ and the $i^{th}$ column of $\bm{V}$. 

In the following simulation studies, all models are fit using Markov Chain Monte Carlo algorithms with Gibbs updates and/or Metropolis-Hastings random-walk updates. Gibbs updates are used for all parameters in the Gaussian models and for $\s$ in the Poisson model.  For the Poisson model, the $\betas$ are updated as in \citet{Hughes} using a random walk with proposal ${\betas}^{[j+1]} \sim N({\betas}^{[j]}, \bm{U}^{-1})$ where $\bm{U}$ is the estimated asymptotic covariance matrix from the non-spatial model, and $\bm{\delta}$ is updated with multivariate random walks with a spherical normal proposal. For all models, where applicable,  $\as=\at = .01$ and $\bs = \bt = 100$. For the Gaussian models, $\betas$ has a flat prior and for the Poisson model $\betas$ has a normal prior with standard deviation of 1000.

To ensure that the Monte Carlo standard errors were sufficiently small, trial simulations were run until a sample path length was found that ensured all models had Monte Carlo standard error $< .01$. For the Gaussian models, a sample path of $80,000$ was sufficient; while for the Poisson models, a sample path of $1,000,000$ was sufficient. Monte Carlo standard errors were calculated using batch means \citep{flegal2008markov,flegal2012mcmcse}.

\subsection{Simulation 1: Gaussian Model} \label{actsim1}
For this set of simulations we consider Gaussian models of the form \eqref{second} for the graph of the 48 contiguous states. We generate data from three models. All  generating models are of the form $\Y = \bm{1} \beta_0  +  \bm{X}_1 \betax + \bm{\nu} + \bm{\epsilon}$ where $\bm{\epsilon} \sim N(\bm{0}, \bm{I})$, $\beta_0 = 1, ~ \betax=2,$ and $\bm{X}_1$ is  randomly generated to be correlated with $.2n$ of the eigenvectors associated with the lowest non-zero eigenvalues. The vector $\bm{\nu}$ is one of three options:  $\bm{0}$ (i.e., the model is a non-spatial linear model), a realization from $\bm{L} \bm{\delta}$  with  $\bm{\delta} \sim N (\bm{0}, \bm{L}^T \bm{Q} \bm{L})$ (the RHZ model \eqref{RHZ}), or a realization from the ICAR prior with $\s =1$ (realizations from the ICAR random effect are generated using Algorithm 2.6 in \citet{Rue}). For each of the choices of $\bm{\nu}$ we simulated 1000 realizations of $\Y$ and fit the NS model, RHZ model, and the ICAR model.

\cref{sim1} lists these coverage probabilities based on 95\% credible intervals with equal tail weights. 
98\% of the credible intervals from an analysis using the RHZ model were nested within the corresponding credible intervals derived from the NS model. To explore how the RHZ analysis model differs from the NS analysis model, we consider how the coverage differed for each data set in \cref{sim1c}. We consider three outcomes: 
\begin{enumerate}
	\item The credible intervals for both models either both contained the true value of $\betax$ or both failed to contain it (Agree) 
	\item  the credible interval of the RHZ model contained the true value of $\betax$ and the NS model did not (RHZ +)
	\item The credible interval of the NS model contained the true value of $\betax$ and the RHZ model did not (NS +).
\end{enumerate}

With respect to coverage, both the ICAR model and the NS model obtain higher coverage rates than the RHZ model, even when the data is generated from the RHZ model. When the data is generated from a NS or RHZ model, the differences between the three models are not substantial. These differences are driven primarily by the widths of the credible intervals. The ICAR model tends to have the largest width and the RHZ model tends to have the narrowest credible intervals. 

Arguably, the setting of primary interest for this and the simulation studies that follow is when the data is generated with an ICAR prior. This setting is often thought to be most closely related to what would be seen in the real world.  When the generating model is the ICAR model, both the RHZ model and the NS model achieve their lowest coverage rates.  This suggests, as observed in \citet{Prates} and \citet{Hanks}, that the RHZ model may suffer poor coverage in the presence of spatial random effects which do not operate orthogonally to the fixed effects. In investigating the cause of this poor performance, we found that it was a combination of bias and narrow credible intervals. Here, we define bias as $\betax$ - $E[\betax | \Y]$. In \cref{appCsim1}, we find that the bias for the point estimates for the RHZ and NS models is at its highest when the generating model is an ICAR model. Since by \cref{thm:Thm1}, the point estimates of the RHZ and NS models are the same, the further reduction in coverage rates for the RHZ are caused by narrower credible intervals. There is not a single case in which the credible interval of the RHZ model obtained coverage when the NS model did not.

\parbox{\textwidth}{
\begin{center}
	\captionsetup{position=above}
	\captionof{table}{Coverage of $\betax$} \label{sim1}
	\nopagebreak
	\begin{tabular}{|p{3cm}|p{3cm}|p{3cm}|p{3cm}|}
		\hline
		\textbf{Analysis Model}  & \multicolumn{3}{|c|}{\textbf{Generating Model}} \\ \hline
		& NS & RHZ & ICAR \\ \hline
		NS & 94.4\% & 98.5\% & 84.6\% \\
		\hline
		RHZ & 93.1\% & 93.8\%  & 72.5\% \\
		\hline 
		ICAR & 97.0\% & 99.4\% & 95.3\% \\
		\hline
	\end{tabular}\\[.3in]
\end{center}}

\parbox{\textwidth}{
\begin{center}
	\captionsetup{position=above}
	\captionof{table}{Comparison of Coverage of the RHZ and NS models} \label{sim1c}
	\begin{tabular}{|p{3cm}|p{3cm}|p{3cm}|p{3cm}|}
		\hline
		\textbf{Comparisons}  & \multicolumn{3}{|c|}{\textbf{Generating Model}} \\ \hline
		& NS & RHZ  & ICAR \\ \hline
		Agree & 98.7\% & 95.3\% & 87.9\% \\
		\hline
		RHZ + & 0.0\% & 0.0\%  & 0.0\% \\
		\hline 
		NS + & 1.3\% & 4.7\% & 12.1\% \\
		\hline
	\end{tabular}\\[.3in]
\end{center}}

\subsection{Simulation 2: Type-S Error in the Gaussian Model} \label{sim2}
For this set of simulations we consider Gaussian models of the form \eqref{second} for the graph of the 48 contiguous states. We again generate data from three models. All models are of the form $\Y = \bm{1} \beta_0 +  \bm{X}_1 \betax_1 + \bm{X}_2 \betax_2 + \bm{\nu} + \bm{\epsilon}$ where $\bm{\epsilon} \sim N(\bm{0}, \bm{I})$, $\beta_0 = 1, ~ \betax_1=2,$ and $\betax_2=0$.  $\bm{X}_1$ is again randomly generated by the process described above to be correlated with $.2n$ of the eigenvectors associated with the lowest non-zero eigenvalues, and $\bm{X}_2$ is independently generated to be correlated with the $.5n$ eigenvectors with the lowest non-zero eigenvalues. Once again, $\bm{\nu}$ is one of the three options listed in Simulation 1.  

With respect to inference on $\betas_1$, we again consider coverage rates. For inference on $\betax_2$ we now consider the Type-S error rate: the proportion of times a credible interval does not contain zero. As in Simulation 1, we use 95\% equal-tailed credible intervals. We are interested in Type-S errors in the context of data which have spatial variation unexplained by the fixed effects. When the generating model is the RHZ or the ICAR model, $\bm{\nu}$ itself provides such variation. In the context of a non-spatial model, the only way to achieve such dependence is by omitting a spatially varying covariate. Thus, when the generating model is the RHZ or ICAR, we use a design matrix of $\X^* = [ \bm{1} ~ \bm{X}_1 ~ \bm{X}_2 ]$. When the generating model is the non-spatial model, we use $\X^* = [ \bm{1} ~  \bm{X}_2 ]$, omitting the spatially varying $\bm{X}_1$. Recall that the associated $\bm{X}$ will always be used when fitting the ICAR model.

The results are contained in \cref{sim2a}- \cref{sim2d}. Both the ICAR model and the NS model have higher coverage rates than the RHZ model in every scenario. As before, there is not a single synthetic data set in which the resulting credible interval obtained coverage for $\betax_1$ under the RHZ model and the NS model did not. The RHZ model once again suffers from poor coverage when the generating model is the ICAR model. The differences in the coverage rates are driven by the same factors as discussed in \cref{actsim1}.

\parbox{\textwidth}{
\begin{center}
	\captionsetup{position=above}
	\captionof{table}{Coverage of $\betax_1$} \label{sim2a}
	\begin{tabular}{|p{3cm}|p{3cm}|p{3cm}|}
		\hline
		\textbf{Analysis Model}  & \multicolumn{2}{|c|}{\textbf{Generating Model}} \\ \hline
		&  RHZ & ICAR \\ \hline
		NS &  97.9\% & 85.3\% \\
		\hline
		RHZ &  90.9\%  & 72.9\% \\
		\hline 
		ICAR &  99.0\% & 95.8\% \\
		\hline
	\end{tabular}\\[.3in]
\end{center}}

\parbox{\textwidth}{
\begin{center}
	\captionsetup{position=above}
	\captionof{table}{Comparison of $\betax_1$ coverage of the RHZ and NS models} \label{sim2c}
	\begin{tabular}{|p{3cm}|p{3cm}|p{3cm}|}
		\hline
		\textbf{Comparisons}  & \multicolumn{2}{|c|}{\textbf{Generating Model}} \\ \hline
		& RHZ  & ICAR \\ \hline
		Agree & 93.0\% & 87.6\%  \\
		\hline 
		RHZ + & 0.0\% & 0.0\% \\
		\hline
		NS +  &  7.0\%  & 12.4\% \\
		\hline
	\end{tabular}\\[.3in]
\end{center}}

In \cref{sim2b}, the RHZ model has extremely inflated Type-S error rates - even when the data is generated from a RHZ model. Notably, the RHZ model performs the worst with respect to Type-S errors when a spatially varying covariate is omitted from the design matrix. In \cref{appCsim2}, we find that the bias for the NS and RHZ point estimates is at its highest in this setting. This bias leads to the elevated Type-S error rates for the NS model. The extreme inflation from the RHZ model relative to the NS model is caused by  narrow credible intervals. We also compare the performance of the RHZ and NS analysis models for each synthetic data set with respect to Type-S error in \cref{sim2d}. Every data set for which the NS model resulted in a Type-S error, the RHZ model did as well. However, there were a a number of data sets for which the RHZ model resulted in a Type-S error, and the NS model did not.
\parbox{\textwidth}{\begin{center}
	\captionsetup{position=above}
	\captionof{table}{Type-S Error of $\betax_2$} \label{sim2b}
	\begin{tabular}{|p{3cm}|p{3cm}|p{3cm}|p{3cm}|}
		\hline
		\textbf{Analysis Model}  & \multicolumn{3}{|c|}{\textbf{Generating Model}} \\ \hline
		& NS & RHZ & ICAR \\ \hline
		NS & 13.9\% & 8.5\% & 10.3\% \\
		\hline
		RHZ & 80.2\% & 18.0\%  & 21.7\% \\
		\hline 
		ICAR & 3.7\% & 4.8\% & 6.5\% \\
		\hline
	\end{tabular}\\[.3in]
\end{center}}

\parbox{\textwidth}{\begin{center}
	\captionsetup{position=above}
	\captionof{table}{Comparison of Type-S of the RHZ and NS models} \label{sim2d}
	\begin{tabular}{|p{3cm}|p{3cm}|p{3cm}|p{3cm}|}
		\hline
		\textbf{Comparisons}  & \multicolumn{3}{|c|}{\textbf{Generating Model}} \\ \hline
		& NS & RHZ  & ICAR \\ \hline
		Agree & 33.7\% & 90.5\% & 88.6\% \\
		\hline
		RHZ + & 0.0\% & 0.0\%  & 0.0\% \\
		\hline 
		NS + & 66.3\% & 9.5\% & 11.4\% \\
		\hline
	\end{tabular}\\[.3in]
\end{center}}

\subsection{Simulation Study 3: Investigating Type-S Error in Poisson Model} \label{sim3}

Much of the research in the literature as well as the results of this paper focus on Gaussian models. However, in practice most of the models discussed are typically used for areal count data. It is not altogether clear whether work in the linear case can be translated into results for the non-linear case.  In this set of simulations, we attempt to investigate whether the results for the linear case are relevant for count data.  To our knowledge, this is the first extensive simulation study of count data for RSR methods.

For this set of simulations, we use the graph of 194 Slovenia municipalities considered in \citet{Reich} and \citet{Prates}. We consider  count data generated from a Poisson model of the form  $\Z= \beta_0 +  \bm{X}_1 \betax_1 + \bm{X}_2 \betax_2 + \bm{\nu} $ using the log link function. The coefficients are defined as follows: $\beta_0 =1 , ~ \betax_1 =1$, and $\betax_2 = 0$. We generate 100 datasets for each scenario. The set-up of the simulation study is otherwise the same as in \cref{sim2}. Fitting the RHZ model for the Poisson model is different from that of the Gaussian model. The details of the RHZ model in the non-linear context are given in \cref{RHZdetails}.

\begin{figure}[H] 
	\centering
	\includegraphics[width=0.8\textwidth,height=.4\textheight]{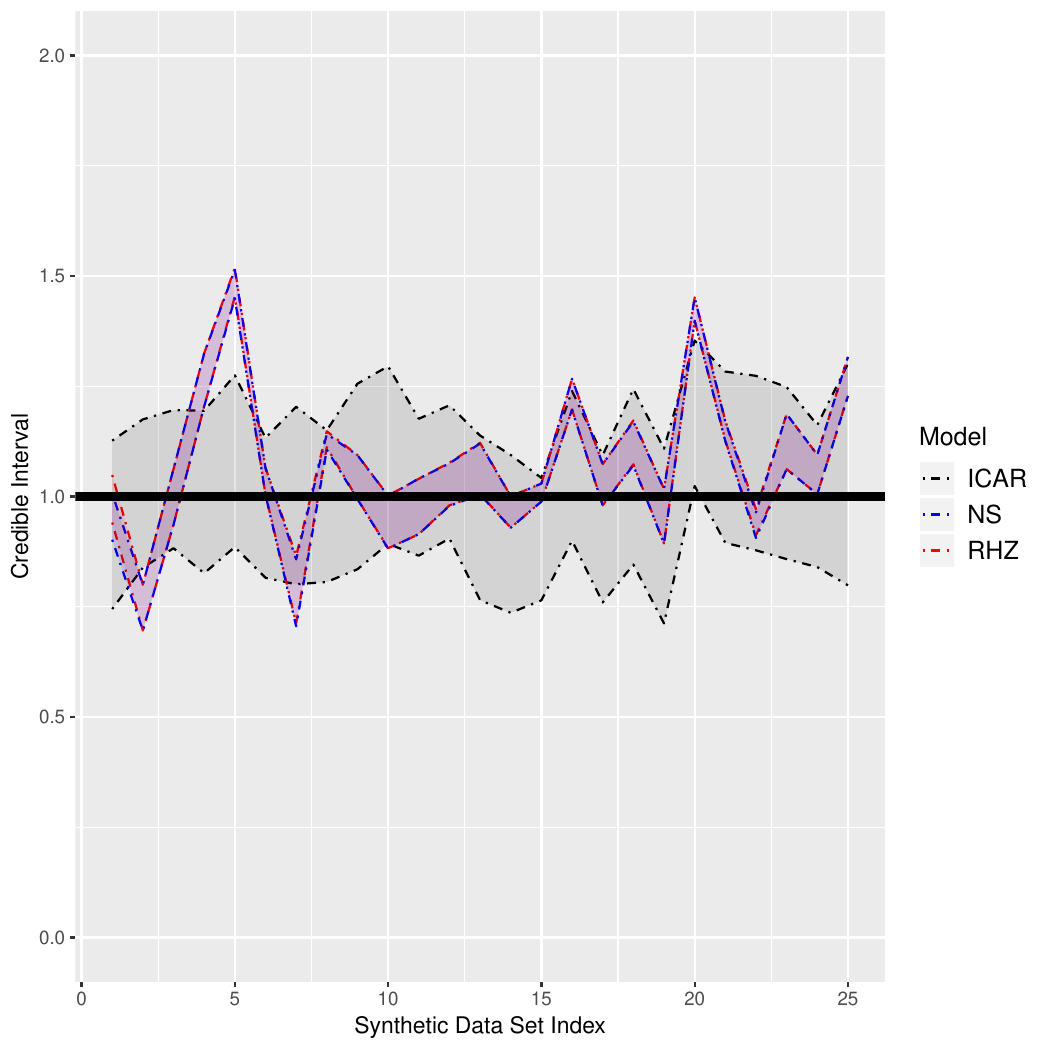}%
	\hfill % <-- Seperation
	\caption[Description] 
	{  Credible intervals for $\betax_1$. This figure was made using the ggplot2 R package \citep{ggplot2}. }
	\label{figcred}
\end{figure}

It should be noted that the credible intervals from the RHZ model are nearly always the same as the credible intervals from the NS model. As an example, \cref{figcred} shows the credible intervals associated with 25 of the 300 simulated datasets. In \cref{sim3a} and \cref{sim3b} we see that that the RHZ model performs essentially the same as the NS model. There are three cases in which the RHZ credible intervals lead to different conclusions than the NS credible intervals. In these cases, the respective bounds for the NS and RHZ credible intervals are within .05 of one another. 

Unlike in the Gaussian case, the performance of the NS model suffers greatly in the presence of spatial variation unexplained by the covariates. This poor coverage seems to be driven by both bias and very narrow credible intervals. In \cref{appCsim3}, we find that the bias for point estimates obtained by the RHZ and NS models is always greater than that of the ICAR model, regardless of the generating model. This phenomena is distinct from the patterns observed in the Gaussian case. With respect to coverage rates and Type-S error, the bias is compounded with very narrow credible intervals for the RHZ and NS models, as illustrated by \cref{sim3b}. Although the Poisson model has not been heavily studied in simulations, we note that \citet{Hughes} observed a similar phenomenon with a single simulated dataset. In that setting, the data was simulated from a HH model. The widths of the 95\% credible intervals obtained from a NS, HH and ICAR models were, respectively, $.24$, $.24$, and $1.84$. It is not clear why the credible intervals of the NS model are so narrow. 

\parbox{\textwidth}{
\begin{center}
	\captionsetup{position=above}
	\captionof{table}{Coverage of $\betax_1$} \label{sim3a}
	\begin{tabular}{|p{3cm}|p{3cm}|p{3cm}|}
		\hline
		\textbf{Analysis Model}  & \multicolumn{2}{|c|}{\textbf{Generating Model}} \\ \hline
		&  RHZ & ICAR \\ \hline
		NS &  32\% & 14\% \\
		\hline
		RHZ &  33\%  & 14\% \\
		\hline 
		ICAR & 97\% & 98\% \\
		\hline
	\end{tabular}\\[.3in]
\end{center}}

\parbox{\textwidth}{
\begin{center}
	\captionsetup{position=above}
	\captionof{table}{Type-S Error of $\betax_2$} \label{sim3b}
	\begin{tabular}{|p{3cm}|p{3cm}|p{3cm}|p{3cm}|}
		\hline
		\textbf{Analysis Model}  & \multicolumn{3}{|c|}{\textbf{Generating Model}} \\ \hline
		& NS & RHZ & ICAR \\ \hline
		NS & 68.0\% & 79.0\% & 67\% \\
		\hline
		RHZ & 67.0\% & 79.0\%  & 67\% \\
		\hline 
		ICAR & 7.0\% & 7.0\% & 3.0\% \\
		\hline
	\end{tabular}\\[.3in]
\end{center}}

\section{Application to Slovenia Data} \label{example}
In light of our findings, we now revisit the dataset which motivated the creation of RSR models. \citet{Reich} and \citet{Hodges2} used the Slovenia stomach cancer data as an example of how spatial confounding could distort inference on regression coefficients when they observed that the 95\% credible interval for the NS model did not contain 0 and the respective credible interval for the ICAR model did contain 0. They believed that this discrepancy occurred because $\textrm{E}[\betass{ICAR} | \Y]$ is biased and $\textrm{Var}[\betass{ICAR} | \Y]$ is overinflated when $\X$ displays spatial dependence. This perspective and the Slovenia dataset has continued to influence subsequent research (See e.g., \citet{Prates}). However, the results of this paper offer an alternative interpretation of why there is a discrepancy between the NS and ICAR models. To explore this, we now introduce the data and corresponding models in more detail.

The data was collected from 1995 to 2001 for each of the 194 municipalities in Slovenia. A full description of the data is available in \citet{hodges2016richly}. The response variable $\Y$ is defined so that $y_i$ is the observed count of stomach cancer cases for the $ith$ municipality. We will let $E_i$ be the expected count of stomach cancer cases and $\X = [x_1, \ldots, x_{194}]$ be the vector of standardized socioeconomic scores for the municipalities \citep[see,][for a description]{Hodges2}. We consider five Poisson models for this model: 1) the non-spatial, 2) the RHZ model, 3) the HH model with all the attractive eigenvectors chosen, 4) the HH model with $q=19$, and 5) the ICAR model. In other words, we assume that the $y_i$ are conditionally independent Poisson random variables with rate parameter $\lambda_i$.  For the non-spatial model, $\log(\lambda_i) = \log(E_i) + \beta_0 + x_i \betax$.  The RHZ model and the HH models are of the form: $\textrm{log}(\lambda_i) = \log(E_i) +\beta_0 + x_i \betax + \left( \W \bm{\delta} \right)_i$. The details of the choices of $\W$ and the prior for $\bm{\delta}$ are given in \cref{RHZdetails}. The ICAR model takes the form $\textrm{log}(\lambda_i) = \log(E_i) + x_i \betax + \bm{\delta}_i$ where $\bm{\delta}$ is assumed to have the ICAR prior. For all models, a normal prior with mean $\bm{0}$ and standard deviation 1,000  is given to the regression coefficients and $\s$ is given a gamma prior with shape and scale, respectively, .01 and 100.

\parbox{\textwidth}{
%\begin{table}[h!]
	\begin{center}
	\captionsetup{position=above}
	\captionof{table}{Summaries of Posterior Distribution $\betax$}
		\label{tab:data}
		\begin{tabular}{l|l|l}
			Model & Posterior Mean & 95\% Credible Interval \\ % <--
			\hline
			NS & -.137 & (-.175, -.098) \\ % <--
			RHZ & -.137 & (-.175, -.098)  \\ % <--
			$\textrm{HH}_{19}$ & -.118 & (-.161, -.074) \\ % <--
			$\textrm{HH}_{\textrm{all pos.}}$ & -.101 & (-.151, -.049) \\ % <--
			ICAR & -.018 & (-.096, .065) \\
		\end{tabular}\\[.3in]
	\end{center}}
%\end{table}}

In \cref{tab:data}, the ICAR model's credible interval is the only one to include zero. In applications, researchers would typically conclude that the socioeconomic scores were not significant for the ICAR model but were significant in the other models. The prevailing belief in spatial confounding literature is that the RSR and NS models offer better inference than the ICAR model for this dataset. However, in \cref{sim3}, we observed that if there is residual spatial dependence (either through a missing spatially varying covariate or through a spatial random effect), the NS and RSR models tend to have more biased point estimates than the ICAR models. Similarly, the NS and RSR models suffer from low coverage rates in this setting. 
Thus, if there is residual spatial dependence unexplained by $\X$, the NS and RSR models will likely offer poor inference for $\betas$.

Unfortunately, the ICAR model also has drawbacks in this setting. Note unlike in \cref{sim3}, the point estimates in \cref{tab:data} are quite small. In this setting, the ICAR model will still have smaller bias and higher rates of coverage than the NS and RHZ models (see \cref{appCsim3a}). However, the Bayesian analogue of power, which we define as percentage of time that the 95\% credible interval does not contain 0, will be very low for the ICAR model (see again \cref{appCsim3a}). Pratically speaking, \cref{thm:Thm1} - \cref{thm:Thm3} suggest that RSR models give essentially the same inference as the NS model in the Gaussian case, and \cref{sim3} seems to support this insight extends to count data. In particular, we note that for each of the RSR models, the 95\% credible intervals are nested within (or equivalent to) the credible interval obtained with the NS model. While it may not be clear when the NS or ICAR model should be preferred, it does seem likely that RSR models will typically be a more computationally intensive method of arriving at the same inference as the NS model with respect to bias and coverage rates.

\section{Discussion}

In this paper we examine the impact that methods designed to alleviate spatial confounding have on inference for regression coefficients of areal SGLMMs.  We emphasis that our inferential focus is limited to assessing whether there is a linear \textit{association} between the response and the covariates in settings where an assumption of (spatial) independence for the errors is apparently invalid (e.g., evidence of residual spatial dependence in the non-spatial model). We note that the distinction between inference related to a linear association and causal inference is often blurred in the literature related to spatial confounding. For instance, the work in \citet{Pactwo} and \citet{Thaden} is arguably more closely related to concerns in causal inference. Teasing apart whether common expectations for regression coefficients in spatial models come from an interest in inference for causal relationships or a linear association will likely be important in future related work. However, the implications of RSR methods for formal causal inference are beyond the scope of the current paper.

The results of \cref{mywork} reexamine the intuition that prompted the development of RSR methods. Despite the expectation that RSR methods will result in marginal posterior variances that are greater than would have been obtained in a non-spatial model, we have shown the opposite is usually true.  Recently, there has been a call for more research to understand when the RSR methods should be used.  This interest is driven in part by the fact that some RSR methods have high rates of Type-S error, particularly when $\frac{\s}{\e}$ is small. Currently, RSR methods are recommended when the covariates $\X$ are correlated with low-frequency eigenvectors of the graph Laplacian. \cref{mywork} and \cref{simu} illustrate how and why the elevated Type-S error rates are occurring for Gaussian RSR models. In particular, these results indicate that it is not the spatial structure of the covariates that drives the performance of RSR models. Rather, the posterior distribution of the regression coefficients for RSR models is primarily a function of how well the spatially varying $\W$ (in RSR models) explains the residual spatial dependence. The question of when the RSR models should be preferred is typically asked with the implicit comparison is to that of a traditional spatial model. Surprisingly, our results indicate that in all cases studied, one would always be better off using a Gaussian non-spatial model than a Gaussian RSR model. For the count data, the RHZ model and non-spatial model result in almost equivalent inference for the regression coefficients and both perform poorly in the presence of spatial variation unexplained by the covariates. This work also shows that the ICAR model can provide poor inference as well, particularly in the context of small effect sizes. 

In spatial statistics, expectations regarding dimension reduction strategies for areal data models have largely been shaped by work in the spatial process modeling realm. For example, in the process model setting, choosing ``attractive" spatial vectors of the Moran operator is analogous to modeling low frequency components of spatial variation. This is the basic premise of approaches such as reduced rank approaches in spatial process modeling. It is quite likely that this method performs well for spatial process modeling. However, this paper shows that insights from spatial process modeling can lead to misleading and counterintuitive results when the interest is no longer on predicting the spatial process. Thus, this work has important implications for dimension reduction when the primary interest is deriving inference about covariate effects. 

\section*{Acknowledgements}
Gabriel Khan is credited for suggesting the clever use of H\"older's Inequality in \cref{thm:Thm1}, which resulted in a relatively cleaner proof. The authors were partially supported by grant NIH NICHD R01-HD088545.  They also acknowledge the support of The Ohio State University's Mathematical Biosciences Institute (NSF DMS-1440386) and the Institute for Population Research (NIH NICHD P2C-HD058484-10).

\pagebreak

\bibliographystyle{apalike} % or try abbrvnat or unsrtnat
\bibliography{Bib} % refers to example.bib

\pagebreak

\appendix

\section{} \label{conditions}

\subsection{Conditions}
\begin{enumerate}
	\item $\F$ is a $q \times q$ non-negative definite and symmetric matrix.
	\item Either $\textrm{rank}(\F) \geq 2$ or $\textrm{rank}(\F) = 1$ and $\as/2 + n/2 - p/2 - q/2 > 1/2$.
	\item $\textrm{rank}(\W) = q \leq n-p$.
	\item $\textrm{rank}(\X) = p < n$.
	\item $\F$ and $ {\W}^T \W $ commute.
\end{enumerate}

\subsection{Details of RHZ and HH Models for Count Data} \label{RHZdetails}

Section 4 of the \citet{Reich} details the construction of a model which smooths orthogonally to the fixed effects for non-linear link functions. In particular, they considered data distributed such that $y_i $ are independent given $\theta_i$ and the probability distribution function is of the form:
\begin{eqnarray*}
	f(y_i | \theta_i )\exp \{ y_i \theta_i  - b(\theta_i) + c(y_i) \}	\\ \nonumber
	g(\mu_i) = \theta_i = \bm{X}_i \betas + (\bm{V} \bm{\delta})_i + \bm{\epsilon}_i % \label{glm}
\end{eqnarray*}
where $E(y_i | \theta_i) = \mu_i$, and  $g$ is a link function. Using this observation and the logic from the Gaussian case, they suggested the following adjustment to address spatial confounding. 
\begin{eqnarray*}
	\bm{\theta} &=&  \X \betas + \hat{\bm{H}}^{-1/2} \bm{L} \bm{\delta}_1 + \hat{\bm{H}}^{-1/2} \bm{L} \bm{\epsilon}_1 \\
	\bm{\delta}_1 &=& \bm{L}^T \hat{\bm{H}}^{1/2} \bm{V} \bm{\delta} \sim N( \bm{0}, \s \bm{L}^T \hat{\bm{H}}^{1/2} \bm{Q} \hat{\bm{H}}^{1/2} \bm{L} )\\
	\bm{\epsilon}_1 &=& \bm{L}^T \hat{\bm{H}}^{1/2} \bm{\epsilon} \sim N(\bm{0}, \e \bm{L}^T  \hat{\bm{H}} \bm{L} )
\end{eqnarray*}
where now $\bm{L}$ are the eigenvectors associated with eigenvalues of 1 from ${\Pperpx}^*$
\begin{eqnarray} \label{genperp}
{\Pperpx}^* = \bm{I} - \hat{\bm{H}}^{1/2} \X (\X^T \hat{\bm{H}} \X)^{-1} \X^T \hat{\bm{H}}^{1/2} 
\end{eqnarray}
and $\hat{\bm{H}}$ is a diagonal matrix with the $(ii)$th element defined to be $ \textrm{Var}(y_i | \betas, \bm{\delta},\bm{\epsilon} )$ evaluated at the mode of $p(\betas, \bm{\delta}, \bm{\epsilon} | \e, \s, \Y)$. In all relevant code, following the choice in \citet{ngspatial}, $\hat{\bm{H}}$ is calculated using the values for the parameters in the last iteration of the Iterated Reweighted Least Squares (IWLS) Algorithm for the non-spatial model. The model detailed in \cref{sim3} is simply a reduced case of this model excluding the heterogeneity effect $\bm{\epsilon}$. In particular the model is:
\[  \Z= \beta_0 +  \bm{X}_1 \betax_1 + \bm{X}_2 \betax_2 + \hat{\bm{H}}^{-1/2} \bm{L} \bm{\delta} \]
\[ \bm{\delta} \sim N( \bm{0}, \s {{\bm{L}}}^T \hat{\bm{H}}^{1/2} \bm{Q} \hat{\bm{H}}^{1/2} {\bm{L}} ) \]

In the application section of \citet{Hughes}, the authors defined the Moran operator to be ${\bm{P}_R}^{\perp} \bm{A} {\bm{P}_R}^{\perp}$ and $\bm{M}_q^*$ to be the eigenvectors of this Moran operator,  where ${\bm{P}_R}^{\perp} = \bm{I} - \bm{R}^{1/2} \X (\X^T \bm{R} \X)^{-1} \X^T \bm{R}^{1/2} $.  In \citet{ngspatial}, the matrix $\bm{R}$ is defined to be a diagonal matrix with $(i,i)$th entry given to be $1/w_i$ where $w_i$ are the weights from the last iteration of the IWLS algorithm in the non-spatial model. Although not explicitly stated it appears that \citet{Hughes} then assumed a model of the following form:
\[  \Z= \beta_0 +  \bm{X}_1 \betax_1 + \bm{X}_2 \betax_2 + \bm{M}^*_q \bm{\delta} \]
\[ \bm{\delta} \sim N( \bm{0}, \s {{\bm{M}}^*}^T  \bm{Q} {{\bm{M}}^*} ) \]

Since this also appears to be the model used in \citet{ngspatial} and studied in \citet{Prates}, we utilize this in \cref{example}.

\section{Proofs of Theorems} \label{proofs}

\subsection{Proof of \cref{thm:Thm1}}

\begin{proof}
	
	\noindent	Under these assumptions, straightforward calculations give the following,
	\begin{flalign*}
	f( \betas | \delta, \s, \e, \Y) &\sim  \textrm{N} \left( ({\X}^T \X)^{-1} {\X}^T \Y ,  (\e {\X}^T \X)^{-1} \right) && 
	%	f( \bm{\delta} | \s, \e, \Y) &\sim& \textrm{N} ( (\frac{\e}{\s} \F + {\W}^T {\W} )^{-1} {\W}^T \Pperpx \Y, \s \F + {\W}^T {\W} \e  )  )
	\end{flalign*}
	where $(\e {\X}^T \X)^{-1}$ is the covariance rather than the precision. This allows us to conclude the following:
	\begin{flalign*}
	\textrm{E} [ \betas |  \Y ] &=  \textrm{E} \left[ \textrm{E} \left[  \betas | \bm{ \delta}, \s, \e , \Y  \right] |  \Y  \right] &&\\ 
	&= \textrm{E} \left[ ({\X}^T \X)^{-1} {\X}^T \Y  |  \Y  \right] && \\
	& =  ({\X}^T \X)^{-1} {\X}^T \Y ,&&
	\end{flalign*}
	\begin{flalign*}
	\textrm{Var} [ \betas |  \Y ] &=  \textrm{Var} \left[ \textrm{E} \left[  \betas | \bm{ \delta}, \s, \e , \Y  \right] |  \Y  \right]  + \textrm{E} \left[ \textrm{Var} \left[  \betas | \bm{ \delta}, \s, \e , \Y  \right] |  \Y  \right] && \\ 
	&= \textrm{E} \left[ (\e {\X}^T \X)^{-1}  |  \Y  \right] && \\
	&= ({\X}^T \X)^{-1} \textrm{E} \left[ \frac{1}{\e} | \Y  \right] &&
	\end{flalign*}

	\noindent The integrals computed above rely on two assumptions:
	
	\begin{enumerate}
		\item $ \Big|  \iiint_{\bm{\delta},\s,\e} f(\bm{\delta}, \s, \e | \Y) d \e d \s d \bm{\delta} \Big| < \infty $
		\item $ \Big| \iiiint_{\bm{\delta}, \bm{\beta}, \s,\e} \frac{1}{\e} f(\bm{\beta},\bm{\delta}, \s, \e | \Y) d \e d \s d \bm{\delta} d \bm{\beta}  \Big| < \infty $
	\end{enumerate}
	
	To show that these conditions both hold, it is sufficient to note the following: 
	\begin{flalign}
	&\Big|  \iiiint_{\bm{\beta}, \bm{\delta},\s,\e} \frac{1}{\e} f(\bm{\beta}, \bm{\delta}, \s, \e | \Y) d \e d \s d \bm{\delta} d \bm{\beta} \Big|  \nonumber \\
	&=  \iiiint_{\bm{\beta}, \bm{\delta},\s,\e} \frac{1}{\e} f(\bm{\beta}, \bm{\delta}, \s, \e | \Y) d \e d \s d \bm{\delta} d \bm{\beta} \nonumber \\ 
	&\propto \iint_{\e, \s} \exp \left \{\frac{-\e}{\bs} \right \} \exp \left \{\frac{-\s}{\bt} \right \} {\s}^{A_{\s}}  \e^{A_{\e}}  \frac{1}{ \sqrt{|   {\W}^T \W  + \frac{\s}{\e } \F | }}   \exp \left \{ \frac{-\e}{2} {\Y}^T \bm{\Sigma}_{\e,\s} \Y \right \}  d \e d \s  \label{bounding}
	\end{flalign}
	where 
	\begin{enumerate}
		\item $\bm{\Sigma}_{\e,\s}  = \Pperpx - {\W} \left( {\W}^T \W + \frac{\s}{\e} \F \right)^{-1} {\W}^T$
		\item $A_{\s} = \textrm{rank}(\F) / 2 + \at -1$
		\item $A_{\e} = \as + n/2 - p/2 - q/2 -2$
	\end{enumerate}
	
	The final integral is finite, which verifies the two assumptions above. To show this, the general strategy is the following.
	The final exponential term can be bounded by a constant using the non-negative definiteness of $\bm{\Sigma_{\e, \s}}$. We first fix $\e$, then for $\s  >1$, convergence occurs due to the exponential decay of the Gamma priors. For $\s<1$, we can ignore the exponential term in $\s$. Doing so, we can use the determinant estimate listed below, Holder's inequality, and Fubini's theorem to directly integrate with respect to $\s$. After applying some standard estimates, we bound the remaining integral with respect to $\e$ by separating it into two cases and bounding each by the sum of independent convergent gamma kernels.
	
	\begin{lemma}[Determinant Estimate] \label{detest}
	Let $\F$ be a non-negative-definite and symmetric $q \times q$ matrix and ${\W}^T \W$ be a positive definite $q \times q$ matrix which commutes with $\F$. Then for $\frac{\s}{\e} >0$ and $j = 1, \ldots, \textrm{rank}(\F)$, we have the following estimate on $\left |{\W}^T \W + \frac{\s}{\e} \F \right | $:
	\begin{eqnarray*}
		\left | {\W}^T \W + \frac{\s}{\e} \F \right | 
		&>&  \left | {\W}^T \W \right | + C_j \left (\frac{\s}{\e} \right )^j. % \\[10pt]
	\end{eqnarray*}
	Here $C_j>0$ is a constant which depends on $j$ (and $\F$).
	
	This lemma can be proven by using the  following observation. Since ${\W}^T \W$ and $\F$ commute, there exists an invertible matrix $\bm{P}$ such that ${\W}^T \W = \bm{P} \bm{\Omega} \bm{P}^{-1}$ and $\F = \bm{P} \bm{\Psi} \bm{P}^{-1}$, where both $\bm{\Omega}$ and $\bm{\Psi}$ are diagonal matrices. Let $\omega_1, \ldots, \omega_q$ be the eigenvalues of ${\W}^T \W$ (i.e., the diagonal of $\bm{\Omega})$ and let $\psi_1, \ldots, \psi_{\textrm{rank}(\F)}$ be the non-zero eigenvalues of $\F$ (i.e., the non-zero diagonal entries of $\bm{\Psi}$). For $j = 1, \ldots, \textrm{rank}(\F)$, we observe the following:
	\begin{eqnarray*}
		\left | {\W}^T \W + \frac{\s}{\e} \F \right |  &=& \prod_{i=1}^{\textrm{rank}(\F)} \left( \omega_i + \frac{\s}{\e}  \psi_i \right ) \prod_{i=\textrm{rank}(\F)+1}^{q}  \omega_i.
	\end{eqnarray*}
\end{lemma}
	
	\begin{lemma}[Non-negative Definiteness of $\bm{\Sigma}_{\e,\s}$] \label{nonnegdef}
		$\bm{\Sigma}_{\e,\s} = \Pperpx -  \W ({\W}^T \W + \frac{\s}{\e} \F)^{-1} {\W}^T$  is non-negative definite.
	\end{lemma}
	This lemma can be proven using the following observations.
	\begin{enumerate}
		\item Because ${\W}^T {\W} $ and $\F$ are simultaneously diagonalizable and ${\W}^T \W$  is positive definite,  $({\W}^T \W)^{-1} - ({\W}^T \W + \lambda \F)^{-1}$ is non-negative definite for all $\lambda \geq 0$.
		\item For any two matrices $\bm{A}$ and $\bm{B}$, if  $\mathcal{C}(\bm{B}) \subset \mathcal{C}(\bm{A}) $, then $||  \bm{P}_{\bm{A}} \Y || ^2 - ||  \bm{P}_{\bm{B}} \Y ||^2 \geq 0$. 
		\item If $q = n-p$, then $\Pperpx = \W ( {\W}^T {\W})^{-1} {\W}^T$. Hence by observation 1,  $ \Pperpx -  \W ({\W}^T \W + \lambda \F)^{-1} {\W}^T $ is non-negative definite for $\lambda \geq 0$.
		\item If $q < n - p$, then $ \Pperpx -  \W ({\W}^T \W + \lambda \F)^{-1} {\W}^T  =   \Pperpx - \bm{P}_{\bm{W}} + \bm{P}_{\bm{W}} -  \W ({\W}^T \W + \lambda \F)^{-1} {\W}^T$. By observations 2 and 3, this is the sum of two non-negative definite matrices and hence non-negative definite.
	\end{enumerate}
	
	With these lemmas in hand, we now return to the problem of showing \eqref{bounding} is finite. 
	Using the non-negative definiteness of $\bm{\Sigma_{\e, \s}}$, we note the following:
	\begin{eqnarray*}
		& & \iint_{\e, \s} \exp \left \{\frac{-\e}{\bs} \right \} \exp \left \{\frac{-\s}{\bt} \right  \} {\s}^{A_{\s}}  \e^{A_{\e}}  \frac{1}{ \sqrt{|   {\W}^T \W  + \frac{\s}{\e } \F | }}   \exp \left \{ \frac{-\e}{2} {\Y}^T \bm{\Sigma}_{\e,\s} \Y \right  \}  d \e d \s  \\
		&\leq &\iint_{\e, \s} \exp \left \{\frac{-\e}{\bs} \right \} \exp \left \{\frac{-\s}{\bt} \right \} {\s}^{A_{\s}}  \e^{A_{\e}}  \frac{1}{ \sqrt{|   {\W}^T \W  + \frac{\s}{\e } \F | }}    d \e d \s 
	\end{eqnarray*}

	\noindent In the following steps, we will use $\lessapprox $ to mean there exists $ D > 0$ such that $\textrm{LHS} \leq D ~  \textrm{RHS}$.
	
	\subparagraph{Case 1: $\s > 1$}
	
	By the \cref{detest} above we can conclude the following
	\begin{eqnarray*}
		& & \int_{1}^{\infty} \int_{0}^{\infty} \exp \left \{\frac{-\e}{\bs} \right \} \exp \left\{\frac{-\s}{\bt} \right \} {\s}^{A_{\s}}  \e^{A_{\e}}  \frac{1}{ \sqrt{|   {\W}^T \W  + \frac{\s}{\e } \F | }}     d \e d \s \\
		& \lessapprox&  \int_{1}^{\infty} \int_{0}^{\infty} \exp \left \{\frac{-\e}{\bs} \right \} \exp \left\{\frac{-\s}{\bt} \right \} {\s}^{A_{\s}}  \e^{A_{\e}}  \frac{1}{ \sqrt{ 1 + \left[\frac{\s}{\e } \right]^{\textrm{rank}(\F)} }}    d \e d \s \\
		&\leq&  \int_{1}^{\infty} \int_{0}^{\infty}  \exp \left \{\frac{-\e}{\bs} \right \} \exp \left\{\frac{-\s}{\bt} \right \} {\s}^{A_{\s}}  \e^{A_{\e}}  \frac{1}{ \sqrt{ \frac{1}{\e^{\textrm{rank}(\F) }  }}}    d \e d \s \\
		&=&   \int_{1}^{\infty} \int_{0}^{\infty}  \exp \left \{\frac{-\e}{\bs} \right \} \exp \left\{\frac{-\s}{\bt} \right \} {\s}^{A_{\s}}  \e^{A_{\e} + \textrm{rank}(\F) /2 }     d \e d \s
	\end{eqnarray*}
	
	Note that the integrand is the product of two gamma kernels: one with shape parameter $A_{\s} +1 =  \textrm{rank}(\F) / 2 + \at > 0$ and scale parameter  $\bt > 0$; the second with shape parameter $A_{\e} + \textrm{rank}(\F)/2 + 1 = \as + n/2 - p/2 - q/2 -1  + \textrm{rank}(\F)/2$ and scale parameter $\bs > 0$. 
	
	\noindent Note that if $\textrm{rank}(\F) \geq 2$, then $A_{\e} + \textrm{rank}(\F)/2 > 0$. If $\textrm{rank}(\F) = 1$ and $\as + n/2 - p/2 - q/2 > 1/2$, then $A_{\e} + \textrm{rank}(\F)/2 > 0$. Thus, under our assumptions this integral is finite.
	
	\subparagraph{Case 2: $\s < 1$}
	
	Again utilizing the determinant estimate:
	\begin{eqnarray*}
		& & \int_{0}^{1} \int_{0}^{\infty}  \exp \left \{\frac{-\e}{\bs} \right \} \exp \left\{\frac{-\s}{\bt} \right \} {\s}^{A_{\s}}  \e^{A_{\e}}  \frac{1}{ \sqrt{|   {\W}^T \W  + \frac{\s}{\e } \F | }}     d \e d \s \\
		& \lessapprox&  \int_{0}^{1} \int_{0}^{\infty}  \exp \left \{\frac{-\e}{\bs} \right \} \exp \left\{\frac{-\s}{\bt} \right \} {\s}^{A_{\s}}  \e^{A_{\e}}  \frac{1}{ \sqrt{ 1 + \left[\frac{\s}{\e } \right]^{\textrm{rank}(\F)} }}    d \e d \s \\
		& < &  \int_{0}^{1} \int_{0}^{\infty} \exp \left \{\frac{-\e}{\bs} \right \}  {\s}^{A_{\s}}  \e^{A_{\e}}  \frac{1}{ \sqrt{ 1 + \left[\frac{\s}{\e } \right]^{\textrm{rank}(\F)} }}    d \e d \s \\
	\end{eqnarray*}
	
	Then using Fubini's theorem, reorganizing terms, and substituting the value of $A_{\s}$ back in:
	\begin{eqnarray*}
		&=&  \int_{0}^{\infty}   \left[  \int_{0}^{1}  {\s}^{\at -1/2}   \frac{{\s}^{\textrm{rank}(\F) /2 - 1/2}}{ \sqrt{ 1 + \left[\frac{\s}{\e } \right]^{\textrm{rank}(\F)} }} d \s \right] \e^{A_{\e}}   \exp \left \{\frac{-\e}{\bs} \right \}    d \e  \\
	\end{eqnarray*}
	
	Then using H\"older's inequality,
	\begin{eqnarray*}
		&\leq&  \int_{0}^{\infty}   \left[  \int_{0}^{1}  {\s}^{2 \at -1} d\s \right]^{1/2}    \left[ \int_{0}^{1} \frac{{\s}^{\textrm{rank}(\F) - 1}}{  1 + \left[\frac{\s}{\e} \right]^{\textrm{rank}(\F)} } d \s \right]^{1/2}   \e^{A_{\e}}   \exp \left \{\frac{-\e}{\bs} \right \}    d \e  \nonumber \\
		&=&  \int_{0}^{\infty}   \frac{1}{\sqrt{2 \at}}    \left[ \frac{\e^{\textrm{rank}(\F) } }{\textrm{rank}(\F) } \log \left( 1 + \frac{1}{\e^{\textrm{rank}(\F) }   }  \right) \right]^{1/2}   \e^{A_{\e}}   \exp \left \{\frac{-\e}{\bs} \right \}    d \e  \nonumber \\
		&\lessapprox&   \int_{0}^{\infty}    \sqrt{ \log \left( 1 + \frac{1}{\e^{\textrm{rank}(\F) }   } \right) } \e^{A_{\e} + \textrm{rank}(\F )/2}   \exp \left \{\frac{-\e}{\bs} \right \}    d \e 
	\end{eqnarray*}
	
	\noindent Now we consider two cases, the first being that $\e < 1$. In this case, note that:
	
	\[ \log \left( 1 + \frac{1}{\e ^{ \textrm{rank}(\F) }}  \right)  \leq \log \left( \frac{2}{\e ^{ \textrm{rank}(\F) }}  \right)= - \textrm{rank}(\F) \log \left( \frac{\e }{2^{\frac{1}{\textrm{rank}(\F)}}}  \right) \]
	
	Note that there exists $C_1(\as) \geq 0$ such that $ - \log \left( \frac{\e }{2^{\frac{1}{\textrm{rank}(\F)}}} \right) < \e^{-\as /2}  + C_1(\as) $  $\forall ~ \e < 1$. Using this inequality,
	\begin{eqnarray*}
		& & \int_{0}^{1}    \sqrt { \log \left( 1 + \frac{1}{\e^{\textrm{rank}(\F) }   } \right)} \e^{A_{\e} + \textrm{rank}(\F) /2 }   \exp \left \{\frac{-\e}{\bs} \right \}    d \e   \\ 
		&\lessapprox&   \int_{0}^{1}   \sqrt{ \e^{-\as /2}  + C_1(\as)}   \e^{A_{\e} + \textrm{rank}(\F)/2}  \exp \left \{\frac{-\e}{\bs} \right \}    d \e \\
		&\lessapprox&   \int_{0}^{1}   \left( \e^{-\as /2}  + 1 \right)  \e^{A_{\e} + \textrm{rank}(\F)/2}  \exp \left \{\frac{-\e}{\bs} \right \}    d \e \\
	\end{eqnarray*}
	
	Note the integrand is the sum of the kernel of two gammas: one with shape parameter $A_{\e} + \textrm{rank}(\F)/2 - \as/2 + 1 = \as/2 + n/2 - p/2 - q/2 +  \textrm{rank}(\F)/2 -1$ and scale parameter $\bs$; the second has shape parameter $A_{\e} + \textrm{rank}(\F)/2 + 1 = \as + n/2 - p/2 - q/2 -1 +  \textrm{rank}(\F)/2$ with the same scale parameter. Under our assumptions, the shape parameters are positive and hence the integral is finite.
	
	If $\e > 1$,  then by the concavity of $\log$, $\log \left( 1 + \frac{1}{\e^{\textrm{rank}(\F) }   } \right) \leq  \frac{1}{\e^{\textrm{rank}(\F) }} $ and:
	\begin{eqnarray*}
		& &  \int_{0}^{\infty}    \sqrt{ \log \left( 1 + \frac{1}{\e^{\textrm{rank}(\F) }   } \right) } \e^{A_{\e} + \textrm{rank}(\F )/2}   \exp \left \{\frac{-\e}{\bs} \right \}    d \e \\
		& \lessapprox & \int_{1}^{\infty}   \sqrt{ \frac{1}{\e^{\textrm{rank}(\F) }} } \e^{A_{\e} + \textrm{rank}(\F) /2 }   \exp \left \{\frac{-\e}{\bs} \right \}    d \e  \\ 
		&=&  \int_{1}^{\infty} \e^{A_{\e} }  \exp \left \{\frac{-\e}{\bs} \right \}  d \e \\ 
		&<&  \int_{1}^{\infty} \e^{A_{\e} +1 }  \exp \left \{\frac{-\e}{\bs} \right \}  d \e 
	\end{eqnarray*}
	
	Note the integrand is the  kernel of a gamma with shape parameter $A_{\e} + 2 = \as + n/2 - p/2 - q/2 > 0$ and scale parameter $\bs$. Thus the integral is finite.
	
\end{proof}

\subsection{Proof of \cref{thm:Thm2}}
\begin{proof}
	Because the marginal posterior variance of the regression coefficients for a model of form \ref{second} under the conditions of \cref{thm:Thm1} is of the form $\textrm{Var}(\betas) = ({\X}^T {\X})^{-1} \textrm{E}(\sigma_{\epsilon}^{RSR} | \Y)$, it is sufficient to show that $\textrm{E}(\sigma_{\epsilon}^{RSR} | \Y) \leq \textrm{E}(\sigma_{\epsilon}^{NS} | \Y)$. In the following proof, we offer a sufficient condition for this inequality to hold. Note that, as in the paper, $r = \frac{\s}{\e}$. 
	
	Basic calculations lead to the following:
	\begin{eqnarray*}
		f(\sigma_{\epsilon}^{NS}| \Y ) &\sim& \textrm{Inv-Gamma}( \as + .5 (n-p), \frac{1}{\bs} + .5 {\Y}^T \Pperpx \Y) \\
		f(\sigma_{\epsilon}^{RSR} | \Y, r ) &\sim& \textrm{Inv-Gamma}( \as + \at + .5 (n-p),  \frac{1}{\bs} + \frac{r}{\bt} + .5 {\Y}^T  (\Pperpx - \W(\bm{I} + r \bm{F}   )^{-1} {\W}^T)   \Y ) \label{tauRHZ}
	\end{eqnarray*}
	
	\noindent Above, we assume a shape-scale parameterization. Note, then, that by properties of the inverse-gamma distribution, we know that:
	\begin{eqnarray*}
	\textrm{E}( \sigma_{\epsilon}^{NS}| \Y) &=& \frac{ \frac{1}{\bs} + .5 {\Y}^T \Pperpx \Y  }{  \as + .5 (n-p) - 1} \\ 
	\textrm{E}(\sigma_{\epsilon}^{RSR} | \Y, r ) &=& \frac{  \frac{1}{\bs} + \frac{r}{\bt} + .5 {\Y}^T  (\Pperpx - \W(\bm{I} + r  \bm{F}   )^{-1} {\W}^T)   \Y}{  \as + \at + .5 (n-p) - 1}
	\end{eqnarray*}
	
	\noindent  By iterated expectations, we can conclude: 
	\begin{eqnarray*}
	\textrm{E}(\sigma_{\epsilon}^{RSR} | \Y ) &=& \frac{  \frac{1}{\bs} + \frac{\textrm{E}(r | \Y)}{\bt} + .5 {\Y}^T  \Pperpx \Y  - \textrm{E} ({\Y}^T \W(\bm{I} + r  \bm{F}   )^{-1} {\W}^T  \Y ) | \Y) }{  \as + \at + .5 (n-p) - 1}
	\end{eqnarray*}
	In using iterated expectations, we rely on the assumption that the moments exist. To see this assumption holds note that it can be shown that $\textrm{E}(r | \Y)$ exists by adding a power of $1$ to $\s$ in the integral in line \cref{bounding} in the proof of \cref{thm:Thm1} and then following the same argument used there.
	We adopt the notation $A =  \textrm{E} ({\Y}^T \W(\bm{I} + r  \bm{F}   )^{-1} {\W}^T  \Y ) | \Y)$ and note that $A \geq 0$. This expectation can be written as the finite sum of terms of $r$ each of which is bounded between 0 and 1, and therefore the integral converges. To show that \cref{thm:Thm2} holds, we then use the following auxiliary function $M(x)$:
	\[ M(x)= \frac{ \frac{1}{\bs} + \textrm{E}(r | \Y) x + .5 \bm{Y}^T \Pperpx \bm{Y} }{\as + kx + .5(n-p)-1}, \]
	where $k > 0$. We consider $M(x)$ on the interval $[0, \frac{1}{\bs}]$. We note that $M(0) = \textrm{E}( \sigma_{\epsilon}^{NS}| \Y) $. For the case that $k = \at \bt$, $M(\frac{1}{\bt}) = \textrm{E}(\sigma_{\epsilon}^{RSR} | \Y ) + \frac{A}{\as + \at + .5 (n-p) - 1}$. As previously noted, $A \geq 0$, and conditions A.1[1]-A.1[4] ensure that $\as + \at + .5 (n-p) - 1 >0$. We now consider the first derivative of $M(x)$:
	\begin{eqnarray*}
	M'(x) &=&  \frac{ \textrm{E}(r | \Y)  \left[ \as + .5(n-p)-1 \right] - k \left[  \frac{1}{\bs}  + .5 \bm{Y}^T \Pperpx \bm{Y} \right] } {\left[ \as + kx + .5(n-p)-1\right]^2 } \\
	&=& \frac{ \left[ \textrm{E}(r | \Y)  - k \textrm{E}( \sigma_{\epsilon}^{NS}| \Y) \right] \left[ \as + .5(n-p)-1 \right] }{\left[ \as + kx + .5(n-p)-1\right]^2 }
	\end{eqnarray*}  
Note that $M(x)$ is monotonically decreasing on $[0, \frac{1}{\bt}]$ whenever:
\[ \frac{E(r | \Y)}{k}  < \textrm{E}( \sigma_{\epsilon}^{NS}| \Y) \]
Now, in the case that $k= \at \bt$, the result follows. As a final note, we observe that $\textrm{E}( \sigma_{\epsilon}^{NS}| \Y)$ is a non-trivial upper bound (i.e., the moment exists) only in the case that $\as + .5(n-p) > 1$.

We offer a final comment that the result will hold under less restrictive assumptions. We focus on a weaker result here because it is more intuitive and relevant to the settings in which RSR models are fit. Intuitively, the more general result suggests that $\textrm{E}(\sigma_{\epsilon}^{RSR} | \Y) \leq \textrm{E}(\sigma_{\epsilon}^{NS} | \Y)$ whenever a spatial precision matrix $\F$ is chosen such that components of $\bm{P}_W \Y$ which are large are not smoothed to zero by the prior on $\bm{\delta}$. In other words, if the spatial basis vectors do a good job of explaining the variability in $\Y$ and the prior on $\bm{\delta}$ does not assume this cannot be true, $\textrm{E}(\sigma_{\epsilon}^{RSR} | \Y) \leq \textrm{E}(\sigma_{\epsilon}^{NS} | \Y)$.  To prove this more general result, we combine the above estimates with a lower bound on the term denoted $A$ using Jensen's inequality. The latter term, in turn, is a function of  the decomposition of $\bm{P}_W \Y$ in terms of the eigenvectors and eigenvalues of $\F$ and the posterior first moment of $r$, which makes the result more technical.

\end{proof}

\subsection{Proof of \cref{thm:Thm3} and \cref{cor:cor1} } 
\begin{proof}
	
	\vspace{.1in}
	
	The general strategy is as follows: we can directly calculate the form of $h(\betax | \Y)$. We find an upper bound for $g(\betax | \Y)$ by bounding an exponential of a non-negative definite matrix by 1, using \cref{detest}, and basic inequalities. Doing so allows us to bound the ratio of $\frac{g(\betax | \Y) }{h(\betax | \Y)}$ by the ratio of polynomials of the same (even) order. The use of L'Hospital's Rule then allows us to draw the conclusions about the tail behavior of the two distributions stated in the theorem.
	
	\noindent \textbf{Form of $h(\betax | \Y)$}
	
	\noindent
	In the univariate case, with the help of Mathematica, it is possible to express $h(\betax | \Y)$ in closed form.
	\begin{eqnarray*}
		h(\betax | \Y) = \frac{1}{D_h} \left[ \frac{ (\Y-\X \betax)^T (\Y - \X \betax)}{2} + \frac{1}{\bs}  \right]^{-\as - n/2} 
	\end{eqnarray*}
	
	where $D_h = \left( \frac{{\X}^T \X }{2}  \right)^{-\as - n/2} \frac{\Gamma(d - 1/2)}{\Gamma(d)}  c^{1/2 -d} \sqrt{\pi} $ is defined such that
	
	\begin{enumerate}
		\item $c = \frac{2}{{\X}^T \X}  \left[ \frac{{\Y}^T \Y}{2} - \frac{ ({\X}^T \Y)^2}{2 {\X}^T \X} + \frac{1}{\bs} \right]$
		\item $d = \as + n/2$
	\end{enumerate}
	
	Note further that:
	\begin{eqnarray}
	h(\betax | \Y)  &=& \frac{ \int f(\betax | \Y, \e) f(\e)  d \e } { \iint f(\betax | \Y, \e) f(\e)  d \e  d\betax  } \nonumber \\ &=& \frac{ (2 \pi)^{-n/2}  \frac{\Gamma(n/2 + \as)}{\Gamma(\as) \bs^{\as}}   \left[ \frac{ (\Y-\X \betax)^T (\Y - \X \betax)}{2} + \frac{1}{\bs}  \right]^{-\as - n/2}   }{   (2 \pi)^{-n/2}  \frac{\Gamma(n/2 + \as)}{\Gamma(\as) \bs^{\as}} D_h}
	\end{eqnarray}
	
	\textbf{Form of $g(\betax | \Y)$}
	\begin{eqnarray}
	g(\betax | \Y)  &=& \frac{ \iiint f(\betax | \Y, \bm{\delta}, \e) f(\bm{\delta} | \s) f(\e) f(\s) d \bm{\delta} d \e d\s } {\iiint f(\betax | \Y, \bm{\delta}, \e) f(\bm{\delta} | \s) f(\e) f(\s) d \bm{\delta} d \e d\s d \betax  } \label{line7}
	\end{eqnarray}
	
	For the moment we will focus on the numerator of \eqref{line7}.
	\begin{flalign*}
	&\iiint f(\betax | \Y, \bm{\delta}, \e) f(\bm{\delta} | \s) f(\e) f(\s) d \bm{\delta} d \e d\s =  &\\
	& \iiint   (2 \pi)^{-n/2 - q/2} {\e}^{n/2} {\s}^{q/2}  f(\s) f(\e)   \exp \left \{ \frac{-\e}{2} (\Y - \X \betax - \W \bm{\delta})^T (\Y - \X \betax - \W \bm{\delta)} \right \} \times & \\
	& \exp \left \{ \frac{-\s}{2} \bm{\delta}^T \F \bm{\delta}   \right \} d \bm{\delta} d \e d \s & 
	\end{flalign*}
	
	We first integrate out $\bm{\delta}$. To do so, with some manipulation, we can recognize a Gaussian kernel and complete the square with:
	\[ \exp \left \{ \frac{\e}{2} ( {\Y}^T \W ({\W}^T \W  + \frac{\s}{\e} \F)^{-1} {\W}^T \Y) \right \}  \]
	
	After which we are left with:
	\begin{flalign*}
	& \iint  (2 \pi)^{-n/2 }  {\e}^{n/2 - q/2} {\s}^{q/2}  f(\s) f(\e) \exp \left \{ \frac{-\e}{2} ({\betax}^T {\X}^T \X \betax - {\betax}^T {\X}^T \Y - {\Y}^T \X \betax) \right \}   \times & \\
	&  \exp \left \{ \frac{-\e}{2} \left( {\Y}^T(\bm{I}- \W ({\W}^T \W  + \frac{\s}{\e} \F)^{-1} {\W}^T) \Y \right) \right \}  \left[ \big |{\W}^T \W + \frac{\s}{\e} \F \big |  \right]^{-1/2}  d \e d \s 
	\end{flalign*}
	
	Note that $\bm{I}- \W ({\W}^T \W  + \frac{\s}{\e} \F)^{-1} {\W}^T$ is non-negative definite. To see this observe that 
	\[ \bm{I}- \W ({\W}^T \W  + \frac{\s}{\e} \F)^{-1} {\W}^T= \bm{I}- \bm{P}_{\W}  + \bm{P}_{\W} - \W ({\W}^T \W  + \frac{\s}{\e} \F)^{-1} {\W}^T. \]
	
	${\W}^T \W$ and $\F$  commute, and we can use the set of observations in the proof of \cref{nonnegdef} to conclude that this is the sum of two non-negative definite matrices and hence non-negative definite. Therefore, we can bound the exponential term with this matrix by 1.
	\begin{flalign*}
	& \leq  \iint  (2 \pi)^{-n/2 }  {\e}^{n/2 - q/2} {\s}^{q/2}  f(\s) f(\e) \exp \left \{ \frac{-\e}{2} ({\betax}^T {\X}^T \X \betax - {\betax}^T {\X}^T \Y - {\Y}^T \X \betax) \right  \}   \times & \\
	&  \left[ \big |{\W}^T \W + \frac{\s}{\e} \F  \big|  \right]^{-1/2}  d \e d \s 
	\end{flalign*}
	
	Again relying on \cref{detest} we can conclude that $\exists ~ C_q > 0 $ such that $|{\W}^T \W + \frac{\s}{\e} \F | > C_q \left (1 + \frac{\s^q}{\e^q} \right)$. Therefore, we have the following bound:
	\begin{flalign*}
	&\leq \iint  (2 \pi)^{-n/2 }  {\e}^{n/2 - q/2} {\s}^{q/2}  f(\s) f(\e)  \exp \left \{ \frac{-\e}{2} ({\betax}^T \X^T \X \betax - {\betax}^T \X^T \Y - \Y^T \X \betax) \right \}   \times & \\
	&   \left[ C_q \left ( 1 + \frac{\s^q}{\e^q}  \right ) \right]^{-1/2}  d \e d \s &\\
	&= \frac{1}{\sqrt{C_q}}  (2 \pi)^{-n/2 } \frac{1}{\Gamma(\as) \bs^{\as}} \frac{1}{\Gamma(\at) \bt^{\at}} \iint   \e^{n/2 + \as -1} \s^{q/2 + \at -1}  \exp \left \{ \frac{-\e}{\bs} \right \}  \exp \left \{ \frac{-\s}{\bt} \right \}  \frac{1}{\sqrt{\e^q + \s^q}}  \times &\\  
	& \exp \left \{ \frac{-\e}{2} ({\betax}^T \X^T \X \betax - {\betax}^T \X^T \Y - \Y^T \X \betax) \right \}   d \e d \s & 
	\end{flalign*}
	
	Now let $m(\e)$ be the pdf of a gamma distribution with shape parameter $n/2 + \as$ and scale parameter $\left[ .5 ({\betas}^T \X^T \X \betax - {\betax}^T \X^T \Y - \Y^T \X \betax) + \frac{1}{\bs}  \right]^{-1}$. Denote $\mu(\betax) = ({\betax}^T \X^T \X \betax - {\betax}^T \X^T \Y - \Y^T \X \betax)$. Substituting in $m(\e)$, we find
	\begin{flalign*}
	&= \frac{1}{\sqrt{C_q}}  (2 \pi)^{-n/2 } \frac{\Gamma(n/2 + \as)}{\Gamma(\as) \bs^{\as}} \frac{\left[ .5 \mu(\betax) + \frac{1}{\bs} \right]^{-\as - n/2} }{\Gamma(\at) \bt^{\at}}  \iint \frac{1}{\sqrt{\e^q + \s^q}} m(\e) \exp \left \{ \frac{-\s}{\bt} \right \} \s^{q/2 + \at -1} d \e d \s &\\
	\end{flalign*}
	
	Note that $ \e^q + \s^q \geq \s^q$, therefore we can construct another upper bound:
	\begin{flalign*}
	&= \frac{1}{\sqrt{C_q}}  (2 \pi)^{-n/2 } \frac{\Gamma(n/2 + \as)}{\Gamma(\as) \bs^{\as}} \frac{\left[ .5 \mu(\betax) + \frac{1}{\bs} \right]^{-\as - n/2} }{\Gamma(\at) \bt^{\at}}  \iint  m(\e) \exp \left \{ \frac{-\s}{\bt} \right \} \s^{ \at -1} d \e d \s &\\
	&= \frac{1}{\sqrt{C_q}}  (2 \pi)^{-n/2 } \frac{\Gamma(n/2 + \as)}{\Gamma(\as) \bs^{\as}} \frac{\left[ .5 \mu(\betax) + \frac{1}{\bs} \right]^{-\as - n/2} }{\Gamma(\at) \bt^{\at}}  \int  \exp \left \{ \frac{-\s}{\bt} \right \} \s^{ \at -1} d \s &\\
	&= \frac{1}{\sqrt{C_q}}  (2 \pi)^{-n/2 } \frac{\Gamma(n/2 + \as)}{\Gamma(\as) \bs^{\as}} \left[ .5 \mu(\betax) + \frac{1}{\bs} \right]^{-\as - n/2}   &
	\end{flalign*}
	
	\noindent \textbf{Considering Ratios of pdfs}
	
	Consider the following, if we let $D_g^* = \iiint f(\betax | \Y, \bm{\delta}, \e) f(\bm{\delta} | \s) f(\e) f(\s) d \bm{\delta} d \e d\s d \betax$
	
	\begin{flalign*}
	& \frac{g(\betax | \Y) }{ h(\betax | \Y)} \leq \frac{\frac{1}{\sqrt{C_q}}  (2 \pi)^{-n/2 } \frac{\Gamma(n/2 + \as)}{\Gamma(\as) \bs^{\as}} \left[ .5 \mu(\betax) + \frac{1}{\bs} \right]^{-\as - n/2} }{(2 \pi)^{-n/2}  \frac{\Gamma(n/2 + \as)}{\Gamma(\as) \bs^{\as}}   \left[ .5(\mu(\betax) + \Y^T \Y) + \frac{1}{\bs}  \right]^{-\as - n/2} } \frac{(2 \pi)^{-n/2}  \frac{\Gamma(n/2 + \as)}{\Gamma(\as) \bs^{\as}} D_h }{D_g^*} & \\
	& =  \frac{\left[ .5 \mu(\betax) + \frac{1}{\bs} \right]^{-\as - n/2} }{ \left[ .5(\mu(\betax) + \Y^T \Y) + \frac{1}{\bs}  \right]^{-\as - n/2} } \frac{(2 \pi)^{-n/2}  \frac{\Gamma(n/2 + \as)}{\Gamma(\as) \bs^{\as}} D_h }{\sqrt{C_q} D_g^*} & \\
	\end{flalign*}
	
	Because $\mu(\betax)$ is a polynomial of even order, this allows us to conclude the following:
	
	\[ \limsup_{\betax \to \infty}  \frac{g(\betax | \Y) }{ h(\betax | \Y)} = \limsup_{\betax \to -\infty}  \frac{g(\betax | \Y) }{ h(\betax | \Y)} \leq  \frac{(2 \pi)^{-n/2}  \frac{\Gamma(n/2 + \as)}{\Gamma(\as) \bs^{\as}} D_h }{\sqrt{C_q} D_g^*}   \]
	
	\noindent \textbf{Implications for Tail Behavior}
	
	Note then by the use of L'Hospital's Rule:
	
	\[ \limsup_{\betax \to \infty}  \frac{1- G(\betax | \Y) }{ 1- H(\betax | \Y)}, ~ \limsup_{\betax \to -\infty}  \frac{G(\betax | \Y) }{ H(\betax | \Y)} \leq  \frac{(2 \pi)^{-n/2}  \frac{\Gamma(n/2 + \as)}{\Gamma(\as) \bs^{\as}} D_h }{\sqrt{C_q} D_g^*}.   \]
	
	Note that if $K(C_q,D_h)= \frac{(2 \pi)^{-n/2}  \frac{\Gamma(n/2 + \as)}{\Gamma(\as) \bs^{\as}} D_h }{\sqrt{C_q} } < D_g^*$, then:
	
	\[ \limsup_{\betax \to \infty}  \frac{1- G(\betax | \Y) }{ 1- H(\betax | \Y)}, ~  \limsup_{\betax \to -\infty}  \frac{G(\betax | \Y) }{ H(\betax | \Y)} < 1 \]
	
\end{proof}

\subsection{Proof of \cref{thm:Thm4}}
\begin{proof}
	
	In the framework of model \eqref{second}, Bayes theorem indicates that the marginal posterior distribution for $\betas$ will be the same for choices of $f, g$ such that:
	\[ f(\Y | \betas, \s, \e) \propto g(\Y | \betas, \s, \e).   \]
	To see this note, that by Bayes Theorem:
	\begin{flalign*}
	f(\betas, \bm{\delta}, \s,\e | \Y) &\propto f(\Y | \betas, \e, \bm{\delta}) f(\bm{\delta} | \s) f(\s) f(\e) f(\betas).
	\end{flalign*}
	So, integrating over $\bm{\delta}$ implies that:
	\begin{flalign*}
	\int f(\betas, \bm{\delta}, \s,\e | \Y) d \bm{\delta}  &\propto \int f(\Y | \betas, \e, \bm{\delta}) f(\bm{\delta} | \s) f(\s) f(\e) f(\betas) d \bm{\delta}.  
	\end{flalign*}
	By the assumption of \textit{a priori} independent prior distributions, then:
	\begin{flalign*}
	f(\betas, \s,\e | \Y)  &\propto f(\s) f(\e) f(\betas)  \int f(\Y | \betas, \e, \bm{\delta}) f(\bm{\delta} | \s)  d \bm{\delta} \\ 
	f(\betas, \s,\e | \Y)  &\propto f(\s) f(\e) f(\betas)  f(\Y | \betas, \s, \e)
	\end{flalign*}  
	
	Hence, substituting  $f(\Y | \betas, \s, \e)$ with  $g(\Y | \betas, \s, \e)$ will yield an equivalent $f(\betas |\Y)$. Straightforward calculations indicate that $f(\Y | \betas, \s, \e)$ will be proportional to:
	\[ \frac{{\e}^{n/2} {\s}^{\textrm{rank}(\F) /2}}{| \e ({\W}^T \W + \frac{\s}{\e} \F)|^{1/2} } \exp \left \{ \frac{-\e}{2} (\Y- \X \betas)^T \left[ \bm{I}- \W [ {\W}^T \W + \frac{\s}{\e} \F  ]^{-1}  {\W}^T \right] (\Y - \X \betas) \right \}	 \]
	
	Assuming that $\X$ and $\Y$ are fixed, this term will be equivalent for the stated choices of ${\W}_i , {\F}_i, i =1,2$. To see this it is sufficient to show the following:
	
	\begin{enumerate}
		\item $\textrm{rank}({\F}_1) = \textrm{rank}({\F}_2)$
		\item $ \left| {\W}_1^T {\W}_1 + \frac{\s}{\e} {\F}_1 \right | =  \left |  {\W}_2^T {\W}_2 + \frac{\s}{\e} {\F}_2 \right |$
		\item ${\W}_1 \left [ {\W}_1^T {\W}_1 + \frac{\s}{\e} {\F}_1  \right ]^{-1}  {\W}_1^T  =  {\W}_2 \left [ {\W}_2^T {\W}_2 + \frac{\s}{\e} {\F}_2 \right ]^{-1}  {\W}_2^T  $
	\end{enumerate}
	
	To see $1$, note: $\textrm{rank}({\F}_1) = \textrm{rank}({\W}_1)  = \textrm{rank}({\W}_2) = \textrm{rank}({\F}_2)$. Conditions 2 and 3 are a consequence of the fact that there exists $\bm{O} \in \textrm{O}(q)$ (i.e., an orthogonal matrix) such that ${\W}_1 = {\W}_2 \bm{O}$ and ${\W}_2 = {\W}_1 \bm{O}^T$. This fact results in the following straightforward observation.
	\begin{eqnarray*}
		\textrm{det}\left ( {\W_1}^T \W + \frac{\s}{\e} {\W}_1^T \bm{B} {\W}_1 \right) &=& \textrm{det} \left ( {\bm{O}}^T {\W}_2^T {\W}_2 \bm{O} +  \frac{\s}{\e} \bm{O}^T {\W}_2^T \bm{B} {\W}_2 \bm{O} \right) \\
		&=& \textrm{det} \left( {\W}_2^T {\W}_2 +  \frac{\s}{\e}  {\W}_2^T \bm{B} {\W}_2  \right ) 
	\end{eqnarray*}
	
	The third condition can be shown with the application of two variants of the Woodbury identity given in lines 156 and 157 of \cite{matrixcookboox}.
	\begin{eqnarray*}
		{\W}_1 \left [ \bm{I} + \frac{\s}{\e} {\W}_1^T \bm{B} {\W}_1 \right ]^{-1} {\W}_1^T &=& {\W}_2 \bm{O} \left [ \bm{I} + \frac{\s}{\e} \bm{O}^T {\W}_2^T \bm{B} {\W}_2 \bm{O}   \right ]^{-1} \bm{O}^T {\W}_2^T \\
		&=&  {\W}_2 \bm{O}\left [ \bm{I} - \bm{O}^T \left ( \bm{I} + \left ( \frac{\s}{\e} {\W}_2^T Q {\W}_2 \right )^{-1} \right)^{-1} \bm{O} \right ] \bm{O}^T {\W}_2^T \\
		&=& {\W}_2 \left [ \bm{I} -\left ( \bm{I} + \left (\frac{\s}{\e} {\W}_2^T Q {\W}_2 \right )^{-1} \right )^{-1}  \right ]  {\W}_2^T \\
		&=& {\W}_2 \left [ \bm{I} + \frac{\s}{\e} {\W}^T \bm{B} {\W}_2 \right ]^{-1} {\W}_2^T.
	\end{eqnarray*}

\end{proof}

\section{Additional Summaries and Simulations}

\subsection{An Overfit NS model for the SAT dataset} \label{overfitexample}

In this subsection, we return to the SAT dataset introduced in \cref{sec32}. We now fit a series of fixed effect models. We start with a model that includes just $\X^*$ (the NS model). We then sequentially add additional synthetic covariates which are columns chosen from an arbitrary basis of $\Pperpx$. To do so, we order (in decreasing value) the columns by their correlation with $\Pperpx \Y$. We add the synthetic covariates in this order to dominate the effect of the $n-p-q$ in the denominator that would be present in \cref{BayesianNS}. We do this until there is insufficient data to update the prior on $\e$ for a final model including 43 synthetic covariates and $\X^*$. As before, $\at= .5, ~ \bt=2000$.

\begin{figure}[H] 
	\centering
	\subcaptionbox{}{\includegraphics[width=0.30\textwidth,height=.22\textheight]{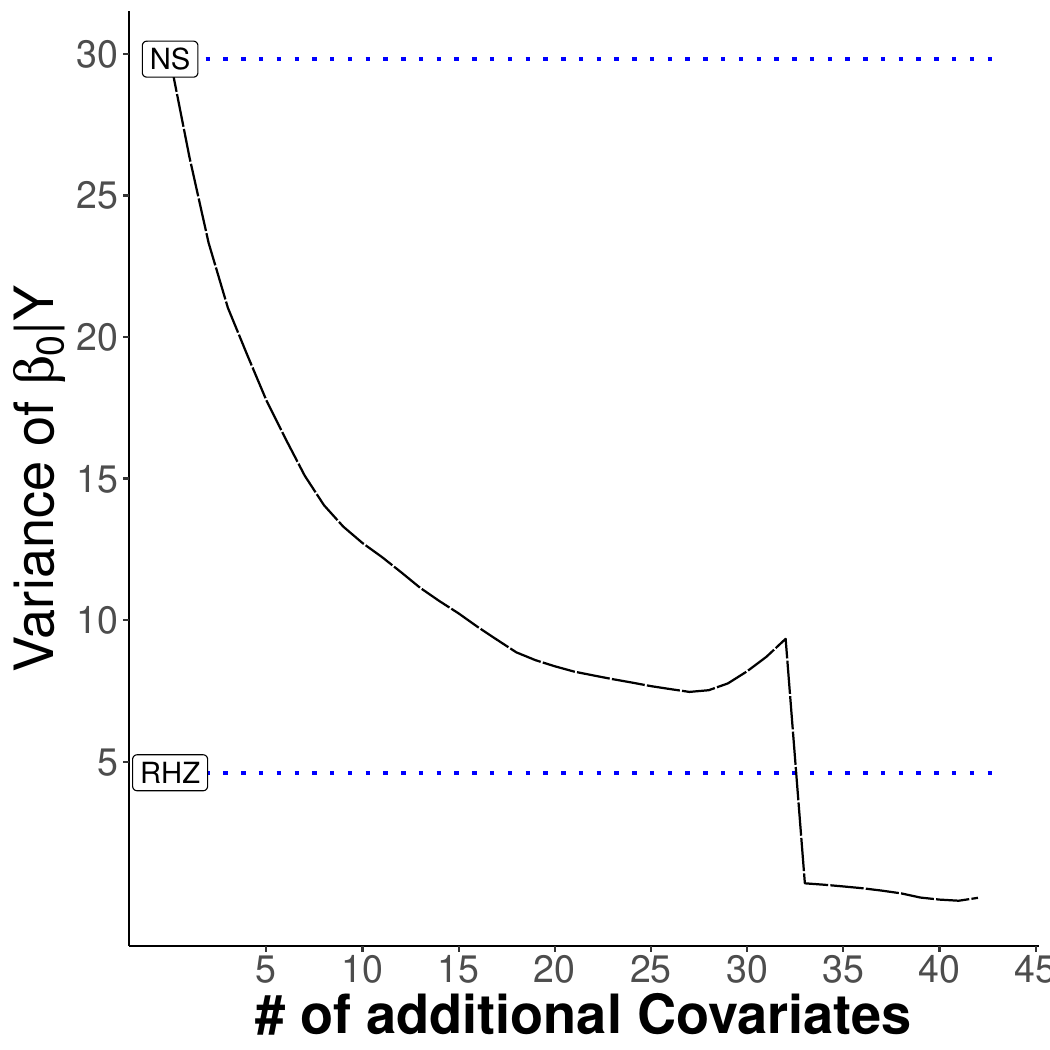}}%
	\hfill % <-- Separation
	\subcaptionbox{}{\includegraphics[width=0.30\textwidth,height=.22\textheight]{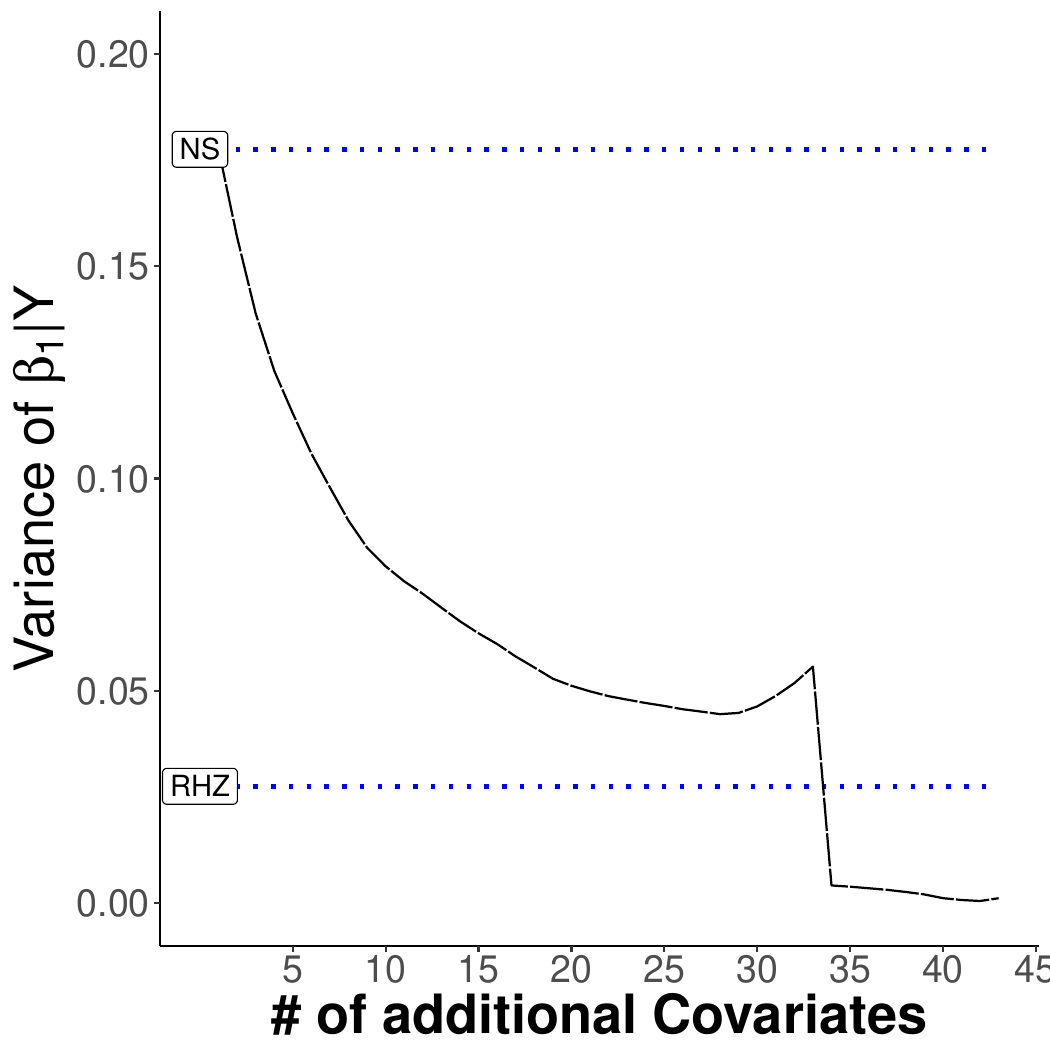}}%
	\hfill % <-- Separation
	\subcaptionbox{}{\includegraphics[width=0.30\textwidth,height=.22\textheight]{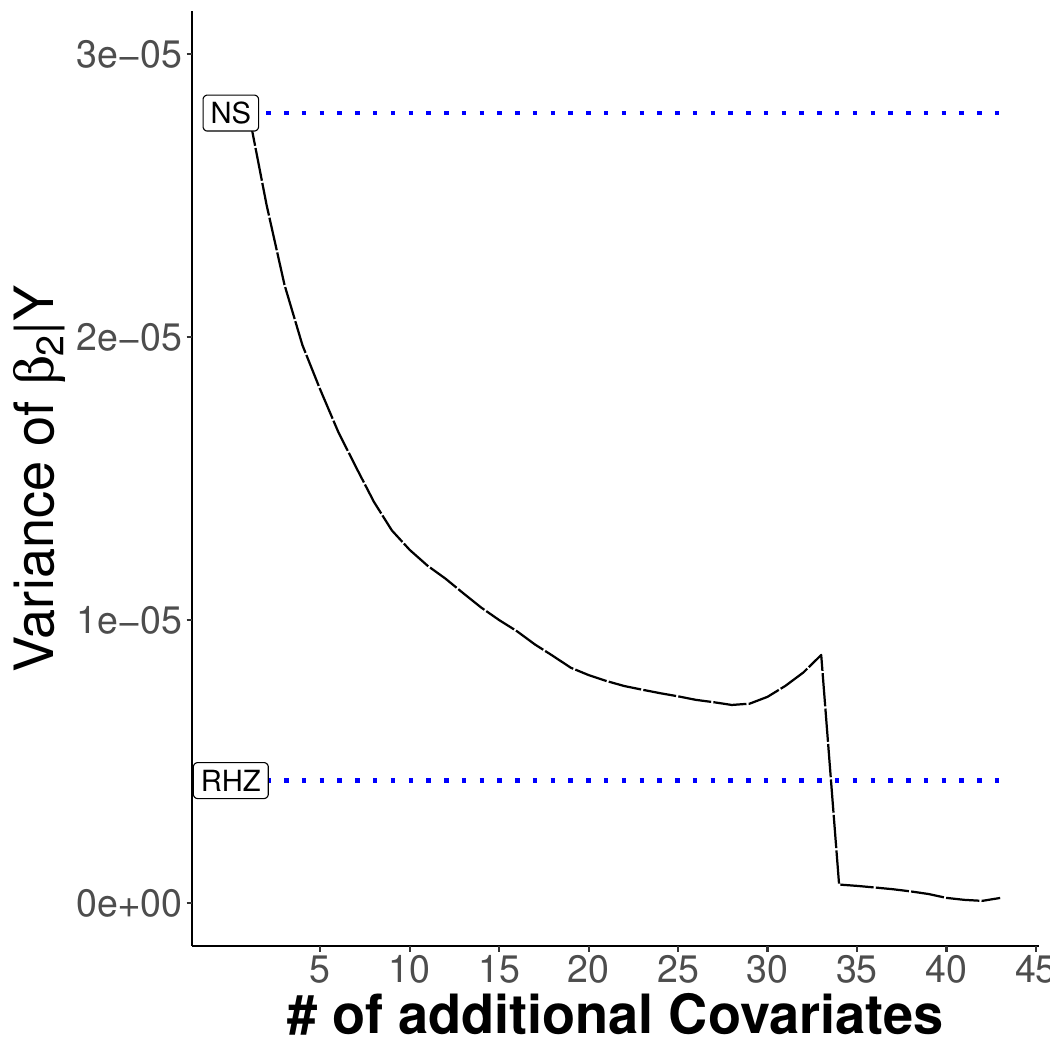}}%
	\caption[Description] 
	{  These graphs depict the posterior variance of each of the regression coefficients as additional vectors from the basis of $\mathcal{C}(\X)^{\perp}$ are added. These figures were constructed using the ggplot2 R package \citep{ggplot2}. }
	\label{fig:maps2}
\end{figure}

In \cref{fig:maps2}, we plot the marginal posterior variance for for $\beta_0$,~ $\beta_1$, and $\beta_2$ for each of the series of fixed effect models. As we expect from the discussion in \cref{sec32}, all models result in approximately the same point estimates of $(590.5, -2.84, .022)$ as additional synthetic covariates and the posterior variances tend to decrease as additional synthetic covariates are added. The relative smoothness of this decrease is due to the fact that we have reduced the impact of the $n-p-q$ term in the denominator of \cref{BayesianNS}. For RSR models, such a tactic is no longer necessary because irregardless of the dimension of $\W$, the corresponding term will be $n-p$.

\subsection{Bias and MSE Results for Simulations 1 to 3} 

In the spatial confounding literature, there is an belief that using the ICAR model results in ``poor" point estimates of the regression coefficients \citep[See e.g.,][]{Reich,Hodges2,Prates}. In looking at the performance of the point estimates, most of the work falls into two groups. First, some work considers the point estimates of the ICAR model to be performing poorly when they differ from the point estimates that would have been obtained in the NS model \citep{Reich,Hodges2}. Other work investigates the poor performance with simulation studies. In this work, poor performance is typically assessed by the difference of the point estimates obtained by the ICAR model and the ``true" (generating) value of the regression coefficients \citep{Hughes}.% We adopt the second approach and define our Bayesian for of "bias" to be $\betax - E[\betax | \Y]$.

%As previously noted, almost all of this work in the Bayesian setting, where bias is not typically of interest.

Almost all of the previous work has focused on the Gaussian case. Although seldom remarked upon, the magnitude of the difference between the point estimates obtained from an ICAR model and the OLS estimates is often very small. For instance, the Slovenia stomach cancer data set has influenced much of the concern with the poor performance of point estimates of the regression coefficients \citep[See e.g.,][]{Reich,Hodges2,Thaden,Prates}. As shown in \cref{example}, the absolute value of the difference between the point estimate obtained in the ICAR model and the point estimate obtained in the NS model is .119. Similar results are found when assessing the performance of point estimates with simulations \citep{Hughes,Prates}. 

In the following subsections, we investigate the performance of the ICAR point estimates in more detail. We find that the difference between the point estimates obtained in the ICAR model and the generating value tends to have small magnitude. We also find that the point estimates of the ICAR model perform better than that of the NS and RHZ models for count data. Although this latter finding contradicts the expectations of the spatial confounding literature, it is not unprecedented. \citet{Hughes} observed something similar in the context of a simulated example.

Although not previously noted, the simulation results for \cref{actsim1}-\cref{sim3} support \cref{thm:Thm2}: in the Gaussian case, the posterior variance for a given regression coefficient for the RHZ model is always less than or equal to that of the posterior variance for the NS model (we considered the results rounded to the 2$nd$ decimal place). As \cref{thm:Thm1} indicates we should expect, the point estimates for regression coefficients for the NS and RHZ model are the same for the Guassian models (we defined same as the absolute difference being less than .01).
Because \cref{thm:Thm1} and \cref{thm:Thm2} were only proven for the Gaussian case, we look to the simulation studies to assess whether similar results hold for the Poisson case. Unlike in the Gaussian case, the marginal posterior means of $\betax$ are not always exactly the same for the NS and RHZ models. However, they are always quite close. The maximum difference is approximately .06 across all simulations. In all but one case, the variance of the marginal posterior distribution for the RHZ model was less than (or equal to) the variance of the marginal posterior distribution for the NS model. This lends support that the findings in \cref{thm:Thm2} may extend in some fort to non-Gaussian cases. The point estimates of the NS and RHZ models are similar enough that bias and mean standard error (MSE) results are the same in almost all cases. Thus, we omit separate categories for the RHZ and NS models in the following discussions of bias and MSE summaries.

\subsubsection{Simulation 1} \label{appCsim1}
In this section, we consider the bias where we define bias as $\betax$ - $E[\betax | \Y]$. To give a sense of the spread of bias in each setting, we calculate the bias for each of the 1000 simulated data sets. \cref{sim1bias} then illustrates the 10$th$ and 90$th$ percentiles of the resulting bias measurements. In \cref{sim1mse}, we consider the average MSE, where the average is taken across the 1,000 simulated data sets.

The results indicate that the bias for the ICAR model tends to be larger than the NS/RHZ model when the data is generated from the RHZ model. If the data is generated with an ICAR random effect, then the bias for the ICAR analysis model tends to be less than the the bias for the NS and RHZ analysis models. The results for the average MSE indicate that there are not large differences in the performance of the various models with respect to this measure. This is not surprising given the relatively small magnitude of the bias for all models. The ICAR analysis model performs marginally better with respect to the MSE than the NS/RHZ model when the data is generated with an ICAR random effect.

\parbox{\textwidth}{
\begin{center}
	\captionsetup{position=above}
	\captionof{table}{10$th$ and 90$th$ Percentiles of Bias for $\betax$} \label{sim1bias}
	\nopagebreak
	\begin{tabular}{|p{3cm}|p{3cm}|p{3cm}|p{3cm}|}
		\hline
		\textbf{Analysis Model}  & \multicolumn{3}{|c|}{\textbf{Generating Model}} \\ \hline
		& NS & RHZ & ICAR \\ \hline
		NS/RHZ & (-.19,.19) & (-.18,.18)  & (-.32,.33) \\
		\hline 
		ICAR & (-.19,.20) & (-.22,.23) & (-.27,.29) \\
		\hline
	\end{tabular}\\[.3in]
\end{center}}

\parbox{\textwidth}{
\begin{center}
	\captionsetup{position=above}
	\captionof{table}{Average MSE of $\betax$} \label{sim1mse}
	\nopagebreak
	\begin{tabular}{|p{3cm}|p{3cm}|p{3cm}|p{3cm}|}
		\hline
		\textbf{Analysis Model}  & \multicolumn{3}{|c|}{\textbf{Generating Model}} \\ \hline
		& NS & RHZ & ICAR \\ \hline
		NS/RHZ & .02 & .02  & .07 \\
		\hline 
		ICAR & .02 & .03 & .05 \\
		\hline
	\end{tabular}\\[.3in]
\end{center}}

\subsubsection{Simulation 2} \label{appCsim2}

We again consider the same summaries of bias and MSE as in \cref{appCsim1}. As before, \cref{sim2bias} shows the NS/RHZ model performs better with respect to bias for $\betax_1$ when the data is generated under an RHZ model. Similarly, the ICAR model performs better when the data is generated with an ICAR random effect. Again, \cref{sim2mse} demonstrates that the average MSEs are relatively small.

Note, that under a RHZ generating model the 10$th$ percentile of the bias differed slightly for the NS (.195) and RHZ (.194) models. However, for convenience we still rounded both to .20 in \cref{sim2bias}.

\parbox{\textwidth}{
\begin{center}
	\captionsetup{position=above}
	\captionof{table}{10$th$ and 90$th$ Percentiles of Bias of $\betax_1$} \label{sim2bias}
	\begin{tabular}{|p{3cm}|p{3cm}|p{3cm}|}
		\hline
		\textbf{Analysis Model}  & \multicolumn{2}{|c|}{\textbf{Generating Model}} \\ \hline
		&  RHZ & ICAR \\ \hline
		NS/RHZ & (-.2,.2) & (-.32,.34) \\
		\hline
		ICAR &  (-.23,.23)  & (-.27, .3) \\
		\hline 
	\end{tabular}\\[.3in]
\end{center}}

\parbox{\textwidth}{
\begin{center}
	\captionsetup{position=above}
	\captionof{table}{Average MSE of $\betax_1$} \label{sim2mse}
	\begin{tabular}{|p{3cm}|p{3cm}|p{3cm}|}
		\hline
		\textbf{Analysis Model}  & \multicolumn{2}{|c|}{\textbf{Generating Model}} \\ \hline
		&  RHZ & ICAR \\ \hline
		NS/RHZ & .02 & .07 \\
		\hline
		ICAR &  .03  & .05 \\
		\hline 
	\end{tabular}\\[.3in]
\end{center}}

%Link this result to CP's comment and above in explaining elevated Type S errors for the NS case
With one exception, the bias and MSE summaries for $\betax_2$ are quite similar to those observed for $\betax_1$, and therefore omitted. The exception occurs when the generating model is the NS model. In this setting the 10$th$ and 90$th$ percentiles for the bias for the NS/RHZ model are $(-.56, .54)$ and the average MSE is .19. The results for the ICAR model remain similar seen for the settings explored in \cref{sim2bias} and \cref{sim2mse}.

\subsubsection{Simulation 3} \label{appCsim3}
In \cref{sim3bias} and \cref{sim3mse}, we consider the same summaries of bias and MSE as in \cref{appCsim2} for $\betax_1$. Note that under the RHZ generating model and with rounding, the 10 $th$ percentile in \cref{sim3bias} is $-.26$ for the NS model and $-.27$ for the RHZ model. For brevity, we just include one case still. %The summaries of bias and MSE are the same for the RHZ and NS model, so we once again omit separate categories. 
Unlike in the Gaussian case, the ICAR model tends to have smaller bias than NS/RHZ models regardless of the generating model. This observation is inconsistent with the general beliefs in the spatial confounding literature \citep[See e.g.,][]{Hodges2}. However, a similar result was seen in an analysis of a single simulated dataset in \citet{Hughes}. 

\parbox{\textwidth}{
\begin{center}
	\captionsetup{position=above}
	\captionof{table}{10$th$ and 90$th$ Percentiles of Bias of $\betax_1$} \label{sim3bias}
	\begin{tabular}{|p{3cm}|p{3cm}|p{3cm}|}
		\hline
		\textbf{Analysis Model}  & \multicolumn{2}{|c|}{\textbf{Generating Model}} \\ \hline
		&  RHZ & ICAR \\ \hline
		NS/RHZ & (-.26,.14) & (-.25,.21) \\
		\hline
		ICAR &  (-.11,.08)  & (-.1, .1) \\
		\hline 
	\end{tabular}\\[.3in]
\end{center}}

\parbox{\textwidth}{
\begin{center}
	\captionsetup{position=above}
	\captionof{table}{Average MSE of $\betax_1$} \label{sim3mse}
	\begin{tabular}{|p{3cm}|p{3cm}|p{3cm}|}
		\hline
		\textbf{Analysis Model}  & \multicolumn{2}{|c|}{\textbf{Generating Model}} \\ \hline
		&  RHZ & ICAR \\ \hline
		NS/RHZ & .04 & .04 \\
		\hline
		ICAR &  .006  & .008 \\
		\hline 
	\end{tabular}\\[.3in]
\end{center}}

The bias and MSE summaries for $\betax_2$ are similar to those observed for $\betax_1$, and therefore omitted.

\subsection{Impact of Small Effect Sizes}

As the previous subsections indicate, the magnitude of the bias is relatively small for all analysis models. In the context of \cref{actsim1}-\cref{sim3}, where the effect sizes are all modest (1 or 2), the bias may not seem terribly concerning. That said, it should be noted that given the very narrow credible intervals of the NS and RHZ models for count data, even a small magnitude of bias can help explain the poor coverage rates and elevated Type-S errors. 

There are many areas of spatial statistics, such as in environmental epidemiology, where small effect sizes are important. Here, to be clear, we define small effect sizes to be associated with regression coefficients of absolute value $\ll 1$. To investigate the impact of spatial confounding in a setting with small effect sizes, we repeat Simulations 1-3. Now, though, we simulate the response ($\Y$ or $\Z$) by replacing $\beta_0 = 1$ and $\betax = \betax_1 =2$ with $\beta_0 = .1$ and $\betax = \betax_1 =.2$. 

For the Gaussian cases, the summaries of bias and MSE typically remain very similar to the results observed in \cref{appCsim1} and \cref{appCsim2}. Given the smaller effect sizes, now the magnitude of the bias is more of a concern. Despite this, the coverage rates observed in \cref{actsim1}-\cref{sim2} remain quite similar. In the following discussions, we omit repeating the coverage, bias, and MSE results for the Gaussian simulations absent a difference from prior results. For both the Gaussian and Poisson models, the results for the Type-S error rates change from \cref{actsim1}-\cref{sim3}. The impact of the smaller effect size can also be seen in the Bayesian analogue of power, which we define as percentage of time that the 95\% credible interval does not contain 0. In \cref{actsim1}-\cref{sim3}, the power was 100\% for all analysis models in every setting. Unsurprisingly, the power suffers with for all models in the context of the smaller effect sizes. We now discuss the results in more detail.

\subsection{Simulation 1} \label{appCsim1a}

\cref{sim1apower} indicates the RHZ model always has the highest power. This difference is most pronounced when the data is generated with an ICAR random effect.  The ICAR model always has the lowest power. Our investigations indicate these differences are explained primarily by the widths of the 95\% credible intervals (as opposed to the impact of bias pushing point estimates away from 0).

\parbox{\textwidth}{
\begin{center}
	\captionsetup{position=above}
	\captionof{table}{Simulation 1: Power for $\betax$} \label{sim1apower}
	\nopagebreak
	\begin{tabular}{|p{3cm}|p{3cm}|p{3cm}|p{3cm}|}
		\hline
		\textbf{Analysis Model}  & \multicolumn{3}{|c|}{\textbf{Generating Model}} \\ \hline
		& NS & RHZ & ICAR \\ \hline
		NS & 27.0\% & 15.1\% & 26.1\% \\
		\hline
		RHZ & 30.7\% & 29.4\%  & 40.0\% \\
		\hline 
		ICAR & 18.6\% & 7.8\% & 10.6\% \\
		\hline
	\end{tabular}\\[.3in]
\end{center}}

\subsection{Simulation 2} \label{appCsim2a}

As in \cref{appCsim1a}, \cref{sim2apower} indicates that the RHZ model always has the highest power and the ICAR model always has the lowest. 

\parbox{\textwidth}{
\begin{center}
	\captionsetup{position=above}
	\captionof{table}{Simulation 2: Power for $\betax_1$} \label{sim2apower}
	\nopagebreak
	\begin{tabular}{|p{3cm}|p{3cm}|p{3cm}|}
		\hline
		\textbf{Analysis Model}  & \multicolumn{2}{|c|}{\textbf{Generating Model}} \\ \hline
		 & RHZ & ICAR \\ \hline
		NS &  14.5\% & 26.8\% \\
		\hline
		RHZ &  29.3\%  & 39.7\% \\
		\hline 
		ICAR &  8.1\% & 11.2\% \\
		\hline
	\end{tabular}\\[.3in]
\end{center}}

\cref{sim2ba} indicates that when the data is generated from a NS model, the Type-S error is much lower than it was in \cref{sim2b} for both the RHZ and NS analysis models. This change is primarily driven by a reduction in the bias for $\betax_2$. For the larger effect sizes, \cref{appCsim2} indicated the 10$th$ and 90$th$ percentiles of the bias were $(-.56, .54)$. Here, the corresponding percentiles are $(-.2,.2)$. The bias for other settings remained largely unchanged.

\parbox{\textwidth}{\begin{center}
	\captionsetup{position=above}
	\captionof{table}{Type-S Error of $\betax_2$} \label{sim2ba}
	\begin{tabular}{|p{3cm}|p{3cm}|p{3cm}|p{3cm}|}
		\hline
		\textbf{Analysis Model}  & \multicolumn{3}{|c|}{\textbf{Generating Model}} \\ \hline
		& NS & RHZ & ICAR \\ \hline
		NS & 5.8\% & 8.5\% & 10.3\% \\
		\hline
		RHZ & 7.9\% & 18.0\%  & 21.7\% \\
		\hline 
		ICAR & 5.0\% & 4.8\% & 6.5\% \\
		\hline
	\end{tabular}\\[.3in]
\end{center}}

\subsection{Simulation 3} \label{appCsim3a}

For the count data, the power depends heavily on how the data was generated. As \cref{sim3apower} indicates, when the data is generated with an IID random variable, all models have exceptionally poor power. The RHZ and NS models have relatively high power (compared to the Gaussian model) when the data is generated with a spatial random effect. 

\parbox{\textwidth}{
\begin{center}
	\captionsetup{position=above}
	\captionof{table}{Simulation 3: Power for $\betax_1$} \label{sim3apower}
	\nopagebreak
	\begin{tabular}{|p{3cm}|p{3cm}|p{3cm}|}
		\hline
		\textbf{Analysis Model}  & \multicolumn{2}{|c|}{\textbf{Generating Model}} \\ \hline
		 & RHZ & ICAR \\ \hline
		NS &  46.0\% & 56.0\% \\
		\hline
		RHZ &   46.0\%  & 57.0\% \\
		\hline 
		ICAR &  13.0\% & 22.0\% \\
		\hline
	\end{tabular}\\[.3in]
\end{center}}

For the count data, the effect size appears to have a more direct impact on the coverage rates, bias, and MSE summaries than it did for the Gaussian data. \cref{sim3aa} indicates that the coverage rates of $\betax_1$ are greatly improved compared to when the generating value of $\betax_1$ was larger (see \cref{sim3a}).

\parbox{\textwidth}{
\begin{center}
	\captionsetup{position=above}
	\captionof{table}{Coverage of $\betax_1$} \label{sim3aa}
	\begin{tabular}{|p{3cm}|p{3cm}|p{3cm}|}
		\hline
		\textbf{Analysis Model}  & \multicolumn{2}{|c|}{\textbf{Generating Model}} \\ \hline
		&  RHZ & ICAR \\ \hline
		NS &  80.0\% & 57.0\% \\
		\hline
		RHZ &  79.0\%  & 58.0\% \\
		\hline 
		ICAR & 100.0\% & 89.0\% \\
		\hline
	\end{tabular}\\[.3in]
\end{center}}

 The driving factor in this reduction appears to be a reduction in the bias for the NS and RHZ models. \cref{sim3biasa} indicates that the bias for the NS and RHZ models have decreased significantly compared to when the data was generated with larger effect sizes (see \cref{sim3bias}). \cref{sim3msea} shows that a similar reduction in the average MSE for these models. The summaries for the bias and MSE measures for $\betax_2$ are once again quite similar to those for $\betax_1$ and thus omitted.
 
 \parbox{\textwidth}{
\begin{center}
	\captionsetup{position=above}
	\captionof{table}{10$th$ and 90$th$ Percentiles of Bias of $\betax_1$} \label{sim3biasa}
	\begin{tabular}{|p{3cm}|p{3cm}|p{3cm}|}
		\hline
		\textbf{Analysis Model}  & \multicolumn{2}{|c|}{\textbf{Generating Model}} \\ \hline
		&  RHZ & ICAR \\ \hline
		NS/RHZ & (-.14,.08) & (-.18,.30) \\ %actually a bit different rhz and ns, but not reporting so didn't worrry about 
		\hline
		ICAR &  (-.11,.09)  & (-.12, .17) \\
		\hline 
	\end{tabular}\\[.3in]
\end{center}}

\parbox{\textwidth}{
\begin{center}
	\captionsetup{position=above}
	\captionof{table}{Average MSE of $\betax_1$} \label{sim3msea}
	\begin{tabular}{|p{3cm}|p{3cm}|p{3cm}|}
		\hline
		\textbf{Analysis Model}  & \multicolumn{2}{|c|}{\textbf{Generating Model}} \\ \hline
		&  RHZ & ICAR \\ \hline
		NS/RHZ & .01 & .03 \\
		\hline
		ICAR &  .006  & .01 \\
		\hline 
	\end{tabular}\\[.3in]
\end{center}}

As with the Gaussian data, a reduction in the bias for the NS/RHZ model corresponds to a reduction in the Type-S error rates, as indicated in \cref{sim3ba}.

\parbox{\textwidth}{
\begin{center}
	\captionsetup{position=above}
	\captionof{table}{Type-S Error of $\betax_2$} \label{sim3ba}
	\begin{tabular}{|p{3cm}|p{3cm}|p{3cm}|p{3cm}|}
		\hline
		\textbf{Analysis Model}  & \multicolumn{3}{|c|}{\textbf{Generating Model}} \\ \hline
		& NS & RHZ & ICAR \\ \hline
		NS & 5.0\% & 33.0\% & 34.0\% \\
		\hline
		RHZ & 5.0\% & 34.0\%  & 33.0\% \\
		\hline 
		ICAR & 4.0\% & 5.0\% & 5.0\% \\
		\hline
	\end{tabular}\\[.3in]
\end{center}}

\end{document}